\documentclass[acmsmall]{acmart}

\usepackage{multirow}
\usepackage{subcaption}
\usepackage{colortbl}
\usepackage{makecell}
\usepackage{pbox}
\usepackage{enumitem}

\usepackage{longtable}
\usepackage{tabularx}
\usepackage{array}
\newcolumntype{Y}{>{\raggedright\arraybackslash}X}

\AtBeginDocument{%
  }

\setcopyright{cc}
\setcctype{by}
\acmJournal{PACMHCI}
\acmYear{2026} \acmVolume{10} \acmNumber{2} \acmArticle{CSCW017}
\acmMonth{4} \acmDOI{10.1145/3788053}

\begin{document}

% \doublespace

\title[``Feeling that I was Collaborating with Them'']{``Feeling that I was Collaborating with Them:'' A 20-year Scoping Review of Social Virtual Reality Leveraging Collaboration}

\author{Niloofar Sayadi}
\affiliation{%
  \institution{University of Notre Dame}
  \city{Notre Dame}
  \country{USA}}
\email{nsayadi2@nd.edu}

\author{Sadie Co}
\affiliation{%
  \institution{University of Notre Dame}
  \city{Notre Dame}
  \country{USA}}
\email{sco@nd.edu}

\author{Diego Gómez-Zará}
\affiliation{%
  \institution{University of Notre Dame}
  \city{Notre Dame}
  \country{USA}
}
\email{dgomezara@nd.edu}




\begin{abstract}
    As more people meet, interact, and socialize online, Social Virtual Reality (VR) emerges as a technology that bridges the gap between traditional face-to-face and online communication. Unlike traditional screen-based applications, Social VR provides immersive, spatial, and three-dimensional social interactions, making it a potential tool for enhancing remote collaborations. Despite the growing interest in Social VR, research on its role in collaboration remains fragmented, calling for a synthesis to identify research gaps and future directions. We conducted a 20-year scoping review, screening 2,035 articles and identifying 62 articles that addressed how Social VR has supported collaboration. Our analysis shows three key levels of support: Social VR can enhance individual perceptions and experiences within their groups, foster team dynamics with virtual elements that enable realistic interactions, and employ the unique affordances of VR to augment users' spaces. We discuss how future research in Social VR should move beyond replicating physical-world interactions and explore how immersive environments can cultivate long-term collaboration, trust, and more diverse and inclusive participation. This review highlights the current practices and challenges, highlighting new opportunities for theorizing and designing Social VR systems that responsibly support remote collaborations.
\end{abstract}

\begin{teaserfigure}
    \centering
    \includegraphics[width=\linewidth]{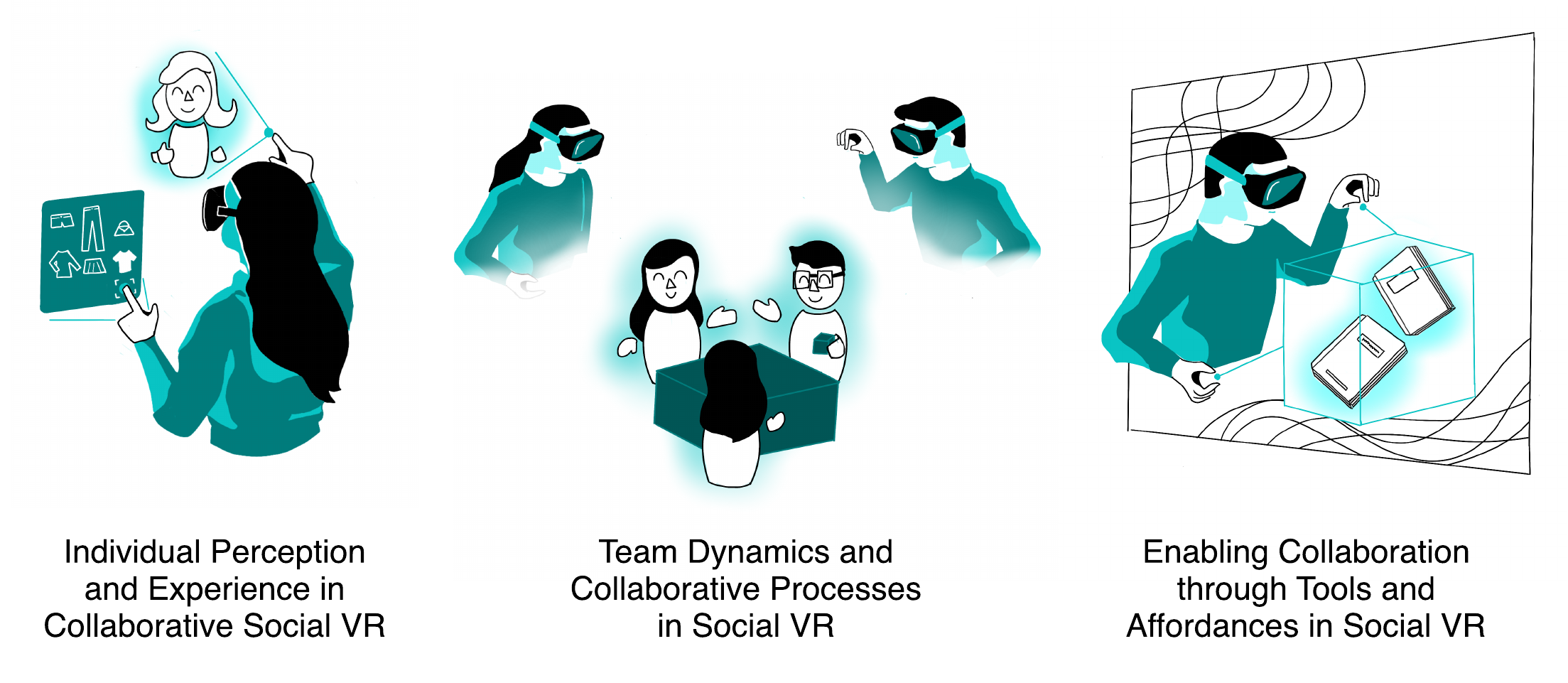}
    \Description{Illustrated overview of a scoping review on Social Virtual Reality collaboration. The figure shows three illustrated themes: individual perception and experience in Social VR, team dynamics and collaborative processes in virtual environments, and enabling collaboration through tools and affordances, depicted through people interacting in immersive virtual settings.}
    \caption{Overview of the scoping review findings on Social Virtual Reality (VR) collaboration. The figure illustrates the three main themes: (1) \textbf{Individual Perception and Experience of Collaboration in Social VR}, examining sense of presence and avatar perception; (2) \textbf{Team Dynamics and Collaboration in Virtual Environments}, focusing on social presence, collaboration realness, trust-building, and engagement; and (3) \textbf{Enabling Collaboration through Social VR Affordances}, addressing virtual environment functionality, virtual objects and tool facilitation, and Supporting Presence through spatial audio and movement.}
    \label{fig:teaser-figure}
\end{teaserfigure}

\maketitle

\section{Introduction}
The rise of Social Virtual Reality (Social VR) is transforming the way people communicate and collaborate in digital spaces. In February 2022, Meta's social VR Horizon Worlds achieved 300,000 active monthly users in the US \cite{Heath2022-dr} and the platform `Rec Room' surpassed 3 million monthly active users \cite{Feltham2022-ex}, showing the growing adoption of Social VR as an environment for social connections. Unlike traditional screen-based communication tools---such as video conferencing or instant messaging---Social VR enables users to immerse themselves in three-dimensional spaces through head-mounted displays (HMDs), creating a sense of physical presence and embodied interaction in simulated environments \cite{maloney2020falling}. It offers unique affordances that can enhance communication among distributed groups, such as avatar embodiment, shared object manipulation, spatial audio, and interactive environments \cite{li2021social, freeman2022working}. Its capacity to simulate real-world experiences has positioned Social VR as a transformative medium for social interactions, particularly after the COVID-19 pandemic, which increased the need for robust alternatives to face-to-face interactions \cite{waizenegger2020affordance,ball2022metaverse,gomez2023promise}.

Despite the growing interest in Social VR, its use in collaborative and task-oriented settings remains unclear. Although most existing research has focused on its applications in gaming, training, and casual gatherings \cite{bujic2021playing,angel2021meta,maloney2020falling}, the potential of Social VR to enable collaborative meetings and teamwork has been underexplored. Companies like Microsoft have attempted to integrate Social VR environments into professional settings to support meetings, conversations, and teamwork \cite{Tilley2022,Roth2023-eo}. However, adoption has often fallen short due to several challenges, including low-quality simulated environments, steep learning curves, cognitive dissonance between physical and virtual worlds, and a lack of applications specifically designed for professional purposes \cite{Ashtari2020,zhao201110,Hubbard2023}. Moreover, the literature has not yet fully explored how these technologies support team processes, such as trust-building, long-term collaboration, and group cohesion \cite{maloney2020falling,abramczuk2023meet,osborne2023being}. These challenges highlight significant gaps in both the design and understanding of Social VR as an environment for distributed collaboration. Specifically, there is a need to better characterize how Social VR environments can address these limitations and replicate the richness of in-person interactions in group settings.

Characterizing studies on Social VR's collaborative uses is crucial for understanding its potential as a transformative tool for distributed collaboration. Research communities in Computer-Supported Cooperative Work (CSCW) and Human-Computer Interaction (HCI) have long sought to design and study systems that support collaboration among remote groups \cite{olson2000distance,harris2019}. As such, Social VR offers an opportunity to rethink how digital tools can enhance collaboration in remote and hybrid contexts, bridging the limitations of current systems while using its unique affordances for presence and engagement. This paper provides a comprehensive overview of the current research landscape in Social VR, identifying critical gaps in areas such as trust-building, user engagement, teamwork scalability, long-term collaboration, the diversity of study samples, and the design of effective applications to support various stages of collaborative processes. By synthesizing these findings, we aim to highlight actionable opportunities for future CSCW research and offer design implications for developing effective collaborative VR systems that address the limitations of existing tools. Given this motivation, the research questions that guide this study are:
\begin{itemize}
    \item \textit{\textbf{RQ1}: How have previous studies addressed team collaboration in Social VR?}
    \item \textit{\textbf{RQ2}: What are the specific gaps, future prospects, and research directions in this area?}
\end{itemize}

To address these research questions, we conducted a scoping review, a methodology designed to map the breadth and diversity of existing research. Unlike systematic reviews, which assess evidence for narrowly defined research questions, our scoping review aims to characterize the landscape and identify solutions and insights that support teamwork in Social VR \cite{munn2018systematic}. This review covered articles published in the IEEE Xplore Digital Library, ACM Digital Library, and Web of Science databases from 2004 to 2024, covering 20 years of research. After screening 2,035 articles, we identified 62 articles that specifically focused on how Social VR has supported teams and enhanced collaboration. Based on this corpus, we conducted a thematic analysis and identified three core themes: (1) \textit{``Individual Perception and Experience of Collaboration in Social VR,''} (2) \textit{``Team Dynamics and Collaboration in Virtual Environments,''} and (3) \textit{``Enabling Collaboration through Social VR Affordances.''} These themes illustrate how Social VR's affordances and mechanisms at the individual, team, and environment levels can facilitate collaboration. 

Our scoping review also highlights several research gaps, including insufficient attention to diverse user populations across demographics and technical expertise, limited understanding of how Social VR scales to larger teams and complex tasks, and a lack of standardized evaluation frameworks for assessing collaboration. Moreover, the themes uncover key questions about trust-building mechanisms, team dynamics, and the transition from traditional online platforms to Social VR for collaboration purposes. Overall, this review provides a comprehensive assessment of the current capabilities and directions that Social VR can offer to individuals and organizations in real-world collaborative settings.

The contributions of this paper are twofold. First, it provides a comprehensive review of how Social VR has been designed and used to support remote collaboration, analyzing both its applications and its impact on teamwork. Second, it provides a holistic road map for advancing Social VR research and system development to strengthen collaboration. Together, the themes and implications discussed in this study provide guidance for CSCW researchers and practitioners seeking to harness the opportunities and address the challenges of making Social VR effective for remote collaboration. 
\section{Background}
We situate our work around prior studies examining (a) teams and technologies to support teamwork, (b) Social VR, and (c) how people have used Social VR to collaborate in task-oriented settings.

\subsection{Teams and Teamwork Technologies}
Teams are groups of two or more individuals working interdependently, adaptively, and dynamically towards a common and valued goal, bringing diverse skills, knowledge, and perspectives to address challenging tasks \cite{gomez2020taxonomy,salas2000teamwork}. Unlike individuals working alone, teams can bring together people with diverse skills and knowledge to achieve better outcomes \cite{engel2015collective,mathieu2017century}. Team members work on different parts of a task, relying on each other, sharing responsibility to achieve a common goal, and resolving complex problems in a distributed manner \cite{mathieu2017century,gomez2020taxonomy,salas2000teamwork}. Previous research shows that successful teamwork relies on clear communication, coordination, mutual support, and individual contributions aligned with the overall task strategy \cite{mathieu2017century}. While teamwork requires more coordination and collaboration than individual efforts, it provides a higher capacity to manage challenging tasks, particularly when specialized skills are needed \cite{mathieu2017century}. The importance of teams has grown as organizational structures have shifted over the past decades from individual roles to team-based systems \cite{wuchty2007increasing}, with tasks increasingly designed around groups working toward shared objectives \cite{arora2023team, mathieu2017century}. For instance, in IT organizations, project teams require programmers, designers, and project managers to work interdependently on complex software development projects, where effective communication, task coordination, and shared understanding are essential for delivering successful outcomes \cite{arora2023team,Huang2023}.

The impact of online technologies on collaboration and teamwork continues to evolve alongside the dynamic technological ecosystem \cite{dulebohn2017virtual,jimenez2017working,treem2020computer}.  Advancements in groupware, intranets, social platforms, smartphones, cloud technologies, and videoconferencing platforms have redefined the degrees of ``virtualness'' \cite{staples2008exploring,brown2020distancing} in which teams operate, ranging from purely face-to-face to only virtual interactions \cite{griffith20018,brown2020distancing}. For example, hybrid workplaces allow team members to work remotely and meet in person on specific occasions \cite{brown2020distancing}. Workers can still be physically present in one building, but coordinate through online communication. Since team members do not necessarily work in the same physical place, current challenges that teams face include coordinating, staying cohesive, and performing effectively \cite{gilson2015virtual}.

With the rise of hybrid and remote work \cite{freeman2022working,bjorn2014does,breideband2023location}, online collaboration tools are now key components in facilitating effective communication and teamwork \cite{cai2019go,li2021social}. Instant messaging platforms, such as Microsoft Teams and Slack, provide another layer of group communication by enabling quick exchanges of information and facilitating asynchronous collaboration \cite{cai2019go,li2021social}. These online communication platforms support various forms of media sharing and integrate with other productivity tools, enhancing workflow efficiency and group awareness. However, they may contribute to information overload and can sometimes lead to misinterpretations without the contextual cues present in verbal communication \cite{cha2018complex}. Video conferencing platforms (e.g., Zoom, Google Meets, Microsoft Teams) also create a shared visual space that enhances user engagement and presence, facilitating real-time communication from different locations \cite{Tang2023,karis2016improving}. Despite offering significant benefits, they often lack the depth of non-verbal cues available in face-to-face interactions, such as body language \cite{freeman2021hugging,bente2008avatar,kanawattanachai2002dynamic}, and can increase mental exhaustion and cognitive load with prolonged use \cite{Bailenson2021-cp}. Previous research has also shown that remote collaboration can curb several activities that require group thinking, such as creativity \cite{brucks2022virtual}, idea generation \cite{lin2023remote}, identity \cite{bartel2007struggle}, and cohesion \cite{yang2022effects}. These current limitations have motivated CSCW researchers to explore how immersive environments, such as Social VR, can improve remote collaboration and enhance the quality of remote teamwork.

\subsection{Social VR}
Social Virtual Reality (VR) refers to ``3D virtual spaces where multiple users can interact with each other using VR head-mounted displays (HMDs)'' \cite{maloney2020falling}. Social VR encompasses a subset of VR platforms and applications specifically designed to support social interaction and co-presence \cite{li2021social,freeman2021body}, where users can feel as if they are physically present with others in the same virtual space \cite{mcveigh2019shaping}. Given these unique affordances, Social VR has the potential to change how people engage, communicate, and collaborate in various activities, from social gatherings and entertainment to educational and professional teamwork settings \cite{deighan2023social}. Platforms like `VRChat' have become popular spaces for individuals to connect in VR environments and foster a sense of community and well-being, particularly with the emergence of the COVID-19 pandemic \cite{osborne2023being}.  

Unlike traditional video conferencing applications, Social VR enables people to immerse themselves in simulated environments, allowing for more interactive ways to interact and socialize \cite{freeman2021hugging}. Social VR applications can enable remote users to interact with 3D virtual objects in immersive spaces \cite{freeman2022working, li2021social}, utilize avatars that can differ from their real bodies and appearances, employ interactive tools that break the physical laws \cite{freeman2022working}, and help people stay more engaged in the team conversation using their bodies and movements \cite{abramczuk2023meet,olaosebikan2022identifying,steinicke2020first}. The unique affordances that VR provides can help users communicate not only with words but also with gestures, body language, and eye gaze with the help of body tracking technologies \cite{freeman2022working}.

Previous research shows that Social VR can enable people to interact in similar ways to face-to-face settings, helping people build stronger social bonds \cite{yoon2021influence}. For entertainment purposes, VR has demonstrated its potential to enhance shared experiences. For example, Lee et al. \cite{lee2022designing} developed a VR lobby for digital opera social experiences, highlighting how Social VR can replicate and enrich live entertainment through immersive environments. Beyond entertainment, Young et al. \cite{young2023case} demonstrated that Social VR can facilitate remote learning by enhancing students' participation in online discussions and activities, improving their sense of presence that traditional online learning platforms often do not provide.
% R&R

% \noindent\textit{Terminology.} We use \emph{immersion} to mean properties of the system (how inclusive, extensive, surrounding, and vivid the display/audio are). We use \emph{presence} to mean the user's feeling of “being there,” shaped by place and plausibility illusions \cite{slater1997framework}. We use \emph{social presence} to mean the feeling of being with other people in VR \cite{biocca2003toward}.
% % R&R
In another study, Sanaei et al. \cite{sanaei2023comparing} conducted a controlled experiment to compare groups working on a circuit repair in face-to-face (F2F), video conferencing, and VR. The study found that participants in VR had a significantly better task experience and lower cognitive efforts compared to participants in the video conferencing condition. These studies suggest that VR can emulate the interactive aspects of in-person interactions that traditional online communication platforms lack. 

Although Social VR has seen increasing use across various settings, its application for remote work and as an online communication tool for group collaboration has not been fully examined.

\subsection{Collaboration Work in Social VR}
Collaboration in Social VR refers to two or more individuals working together in a Social VR application toward a common goal \cite{freeman2022working}. The distinct affordances provided by Social VR, such as avatar customization, nuanced non-verbal behaviors, and spatial interactions, can facilitate a strong sense of being together and foster relationship building among distributed collaborators \cite{freeman2021hugging}. It also enables remote teams to engage in various collaborative tasks that mirror in-person teamwork, from creative projects and problem-solving to mundane tasks and social experiences \cite{freeman2022working}. Furthermore, Social VR allows the interplay of productivity and sociality that often characterizes co-located collaboration \cite{maloney2020falling}. In one study, Maloney et al. \cite{maloney2020falling} conducted interviews to explore what makes activities on Social VR platforms meaningful to their users. They found interactions through full-body tracking and facilitating more natural and expressive communication as the main advantages of Social VR, overcoming the limitations of 2D video conferencing. In another study, Li et al. \cite{li2021social} argue that Social VR affords more natural social interactions than video conferencing, such as the ability for users to organically break off into small groups and interact with virtual objects in the shared environment. Unlike video conferencing, which restricts users to `talking head experiences'---static, front-facing video feeds---Social VR allows for more physical activities and spontaneous collaborations that naturally arise from users' interactions \cite{li2021social}.

Since the COVID-19 pandemic accelerated the shift to remote collaboration, many conference organizers transitioned to online events with mixed success in replicating the traditional experiences, exposing both the strengths and weaknesses of existing conferencing tools \cite{moss2021forging}. Several technology companies and research communities explored metaverse technologies---including Social VR platforms---to support synchronous interactions and socialization within 3D virtual environments \cite{Cheng2022,Duan2021,gomez2023promise}. Platforms like ``Altspace VR'' have demonstrated their ability to foster social presence, enabling remote participants to experience a sense of togetherness within a shared, simulated space \cite{pizzolante2023awe}. Yet, researchers have also pointed out potential challenges, including accessibility issues, privacy concerns, and the need to establish new social norms for virtual interactions occurring in immersive spaces \cite{moss2021forging,gomez2023promise}.

Given the growing research on Social VR, we aim to map how previous research has studied these technologies in the context of collaborations and teamwork. Synthesizing key findings and implications from these studies can reveal the specific benefits of Social VR for team dynamics, inform the design of more effective and safe collaboration applications, and identify barriers to adoption. As technologies continue to reshape how people work together \cite{harris2019}, the CSCW community is well positioned to examine how Social VR's unique capabilities and affordances enable remote collaborations and guide future system development \cite{Wallace2017,Gonzales2015}.

\section{Methodology}
We followed a scoping review methodology \cite{arksey2005scoping} to synthesize existing knowledge and identify trends in using Social VR for collaborative purposes. This methodology allows researchers to map the literature and identify gaps in a specific area of research \cite{arksey2005scoping,levac2010scoping}. Scoping reviews have several advantages: they are transparent, comprehensive, less prone to bias, and make it easier to reproduce the detailed information reported about each step of the review and how it is conducted \cite{pham2014scoping}. They have gained prominence in the HCI and CSCW communities as they enable the synthesis and comprehension of the development of different research directions \cite{Wallace2017,cosio2023virtual,bergram2022digital,rogers2022much,wei2022communication}.

We used the Preferred Reporting Items for Scoping Reviews and Meta-Analyses (PRISMA) guidelines to report the methods and results \cite{page2021prisma}. PRISMA covers four stages in the review process: Identification, Screening, Eligibility, and Inclusion. We collected papers and tabulated them in a shared Google Spreadsheet, capturing metadata such as publication year, abstract, and keywords. Once the metadata was recorded, authors' names were hidden to avoid bias during the coding phase.

Consistent with scoping-review aims, we did not conduct a formal critical appraisal of individual studies; our goal was to map the evidence and characterize study features.

\subsection{Search Strategy and Data Sources}
To identify relevant articles for our scoping review, we developed a comprehensive search query that could capture the key concepts of our study while balancing sensitivity and specificity. After several iterations, we built our search query by breaking down our research topic into its core components: the technology (i.e., social VR), the activity (i.e., collaboration), and the groups (i.e., teams). For the technology concept, we included several synonyms and related terms to maximize coverage, namely ``Social Virtual Reality,'' ``Social VR,'' ``Multi-user Virtual Reality,'' and ``Collaborative Virtual Environments.'' We intentionally did not include the term ``Virtual Reality'' alone to focus on studies that specifically addressed Social VR. For the collaboration concept, we chose a set of keywords focused on people working together: ``Collaborative,'' ``Collaboration,'' and ``Cooperation.'' We chose to add the term ``cooperation'' alongside ``collaboration,'' as the two are closely related and often used interchangeably in the literature \cite{stout2021collaboration}. Lastly, for the groups concept, we included the words referring to teams: ``Team,'' ``Teams,'' ``Teamwork,'' ``Group,'' and ``Groups.'' These words defined the importance of studies analyzing multiple individuals or users within a unit. We combined the synonyms within each concept using the Boolean OR operator and then linked the three main concepts with the AND operator. As a result, our final query was: \textit{(``Social Virtual Reality'' OR ``Social VR'' OR ``Multi-user Virtual Reality'' OR ``Collaborative Virtual Environments'') AND (Team OR Teams OR Teamwork OR Group OR Groups) AND (Collaborative OR Collaboration OR Cooperation).}

We conducted the first step, ``Identification,'' using the following three databases: IEEE Xplore Digital Library, ACM Digital Library, and Web of Science. We chose these datasets since they cover the leading conferences and journals on human-computer interaction and team science. We searched for articles in these databases using their titles, abstracts, and full text. We conducted the final search on February 19, 2025. We included articles published between 2004 and 2024, covering 20 years of research. We limited the included papers to the last 20 years to ensure that our corpus captured the most recent developments, which led to the significant boom in HMD technologies starting around the 2010s, including HTC Vive and Oculus Rift. We also checked earlier studies, which began as early as 1993 in our search scope, and primarily focused on collaborative virtual environments (CVEs) \cite{benford2001collaborative,churchill1998collaborative} using screen-based technologies rather than HMDs. For the few early papers involving VR, their contributions were either revisited in later research or shaped by the technological constraints of the time, which limited their applicability to the current landscape of Social VR research. 

\begin{figure}[!htb]
  \centering
  \includegraphics[width=0.7\textwidth]{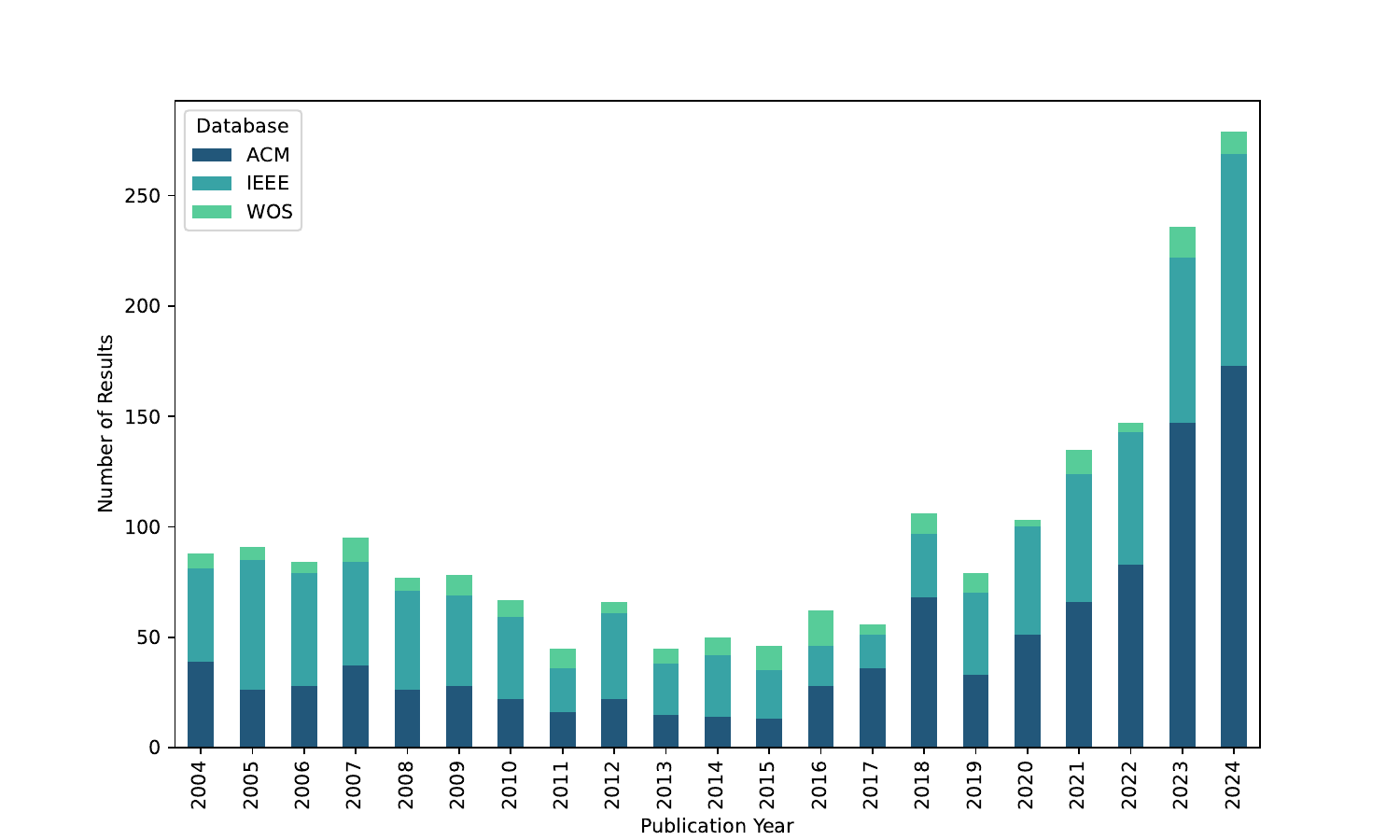}
  \Description{Bar chart showing the number of search results returned by three databases, ACM Digital Library, IEEE Xplore, and Web of Science, for each publication year from 2004 to 2024. The bars indicate the annual volume of papers included in the initial search corpus.}
  \caption{The number of search results of the final query in three databases between 2004-2024.}
  \label{trendofpapers}
\end{figure}

We obtained 892 papers from the IEEE Xplore Digital Library, 972 papers from the ACM Digital Library, and 182 papers from the Web of Science database, totaling 2,046 papers. Since the ACM Digital Library does not allow users to export results from its website, we manually copied and pasted the results into a Google Spreadsheet. For the other two datasets, we exported the results from their websites to a CSV file. We then compiled all the results into a single spreadsheet for analysis. In cases where we found two versions of an article (e.g., a conference paper later extended into a journal article), we included the most recent version. Based on the titles, we removed 11 duplicate articles. This left us with 2,035 articles for screening. Fig.~\ref{trendofpapers} depicts the trend of paper counts for each year in our search corpus from 2004 to 2024.

\subsection{Eligibility Criteria}
\paragraph{Inclusion} We included articles that specifically addressed collaboration within Social VR, focusing on how it can support people and organizations in collaborating and accomplishing work-related tasks. Our inclusion criteria were: 
\begin{itemize}
    \item Articles must involve two or more human users collaborating in Social VR. Collaboration involves communication, coordination, and sharing of knowledge and resources among participants to achieve a joint objective.
    \item Articles must focus on Social VR applications experienced through head-mounted displays (HMDs). 
    \item The collaborative activity or task described in the paper must have a clear goal relevant to a work setting, such as problem-solving, decision-making, or a creative project. Simply meeting, learning, chatting without a shared work-related purpose, or casual socializing were not considered collaborative tasks for this review.
    \item We included only full papers published in journals or conference proceedings to focus on more substantial and in-depth contributions to the literature on collaboration in Social VR.
\end{itemize}

\paragraph{Exclusion} We decided to exclude articles that were not focused on our research questions or were not accessible to the research team. Our exclusion criteria were:
\begin{itemize}
    \item Articles that examined interactions only between virtual agents without any human users. Our interest was in understanding collaboration between real human users.
    \item Articles where the VR environment included only a single human participant interacting with virtual agents were also excluded, as this does not constitute a human team.
    \item Articles that only examined desktop-based virtual environments or other screen setups.
    \item Articles that primarily focused on other forms of extended reality (XR) technologies besides VR, such as augmented reality (AR) and mixed reality (MR). While these XR technologies are related to VR and can also support remote collaboration, they involve different technical setups, interaction modalities, and user experiences. To maintain a clear scope, we limited our review to papers that specifically examined collaboration within fully immersive virtual reality environments using head-mounted displays.
    \item Articles primarily focused on (i) educational settings, such as teaching in classrooms or to children; (ii) training either for medical purposes or for specialized fields; and (iii) gaming, whether for entertainment or training purposes. We excluded these papers to concentrate the review on work contexts.
    \item Workshop papers, magazines, posters, extended abstracts, and thesis papers. While potentially insightful, these formats were often too brief to provide a substantial contribution. 
    \item Articles written in languages other than English were excluded to avoid potential misinterpretation.
    \item Articles that could not be accessed or downloaded from their original published source.
\end{itemize}

\subsection{Article Selection}
\begin{figure}[!htb]
  \centering
  \includegraphics[trim={5cm 0 5cm 0},clip,width=\columnwidth]{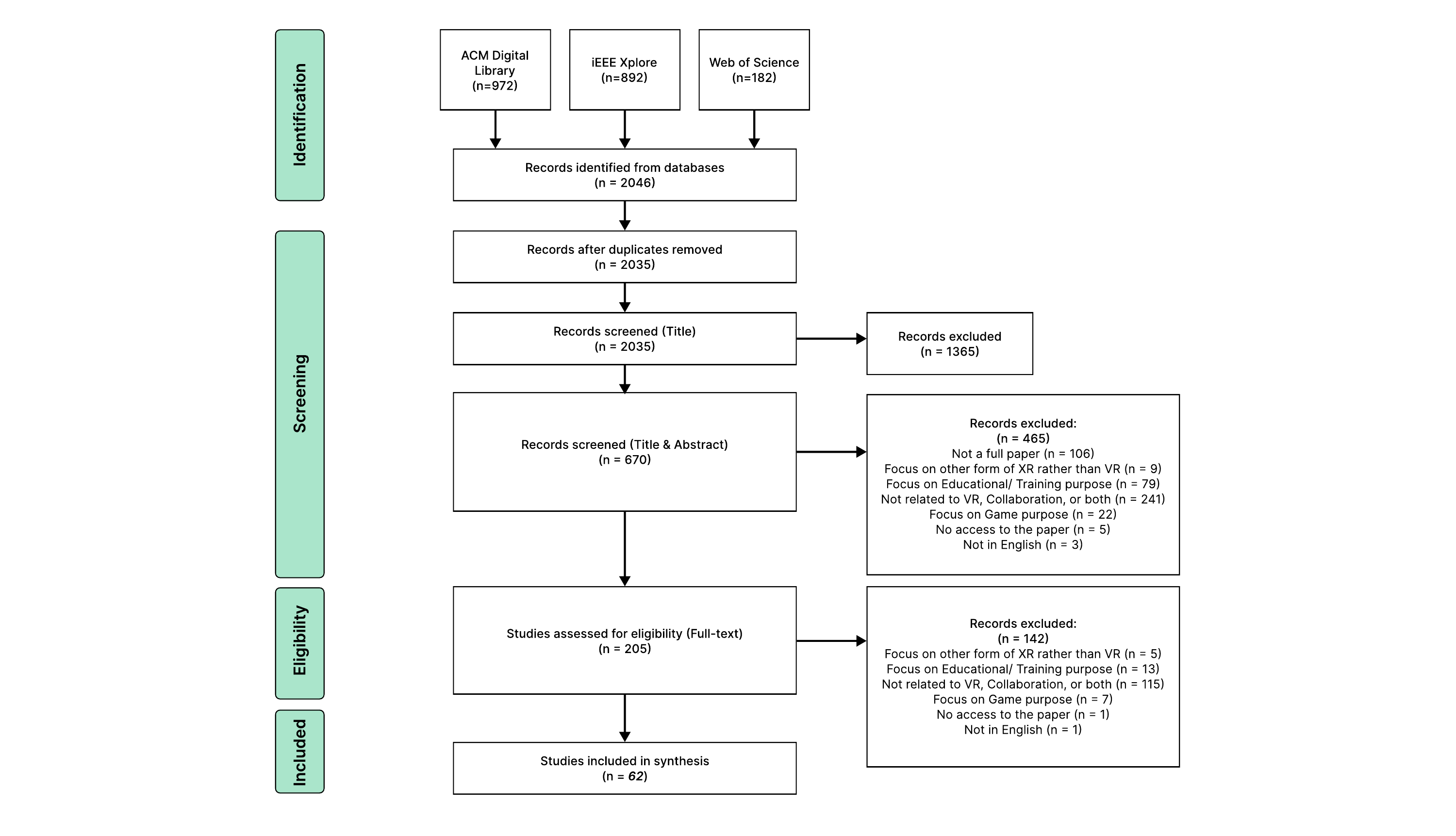}
   \Description{PRISMA flow diagram depicting the article selection process. The figure shows the number of records identified from ACM Digital Library, IEEE Xplore, and Web of Science, followed by stages of duplicate removal, title and abstract screening, full-text eligibility assessment, and the final inclusion of studies in the review.}
  \caption{PRISMA Flow Diagram of this study. It presents the details of the article selection process.}
  \label{prisma}
\end{figure}

\begin{figure}[!htb]
  \centering
  \includegraphics[width=0.7\textwidth]{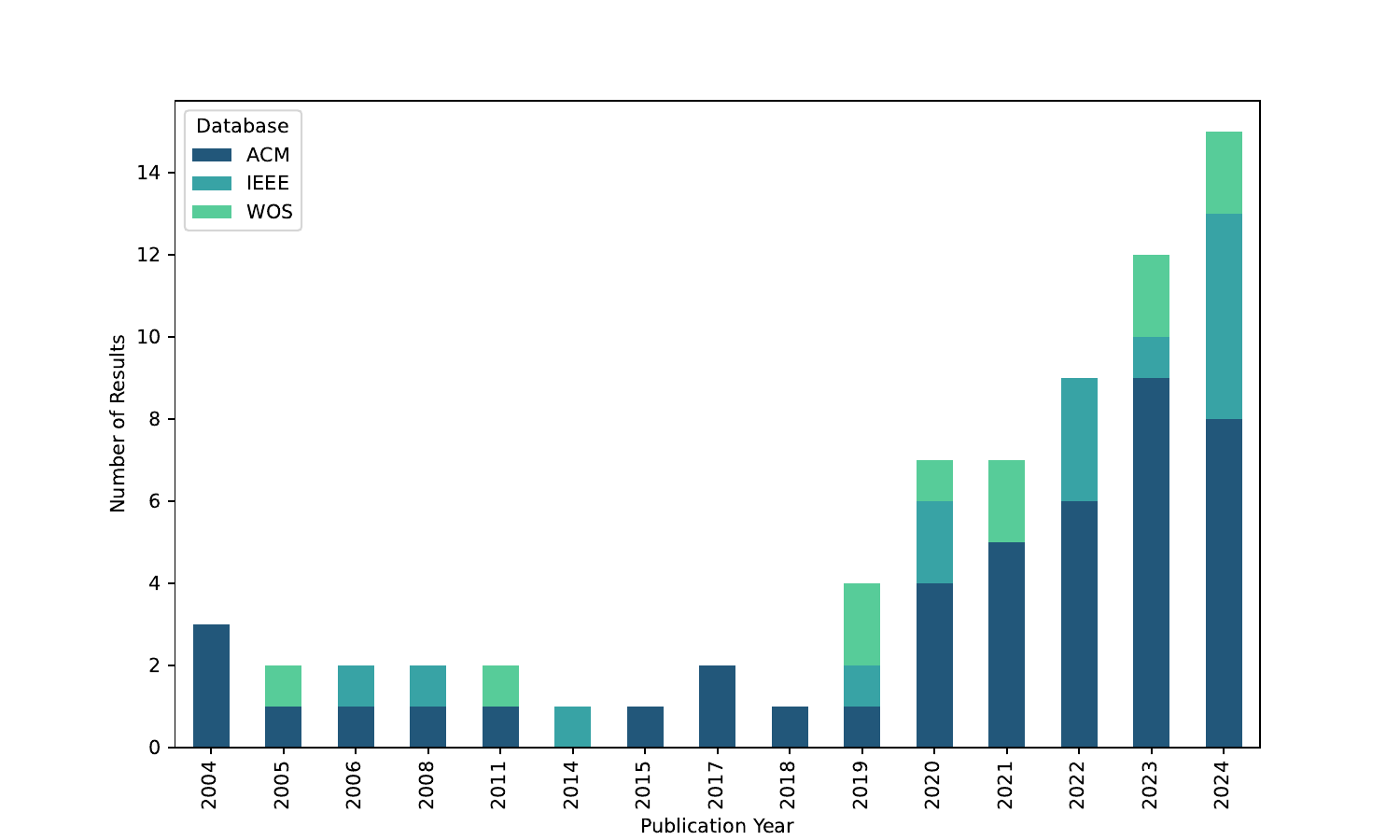}
  \Description{Bar chart showing the yearly distribution of the 62 included articles across three databases, ACM Digital Library, IEEE Xplore, and Web of Science, from 2004 to 2024. Bars represent publication years, with different colors indicating contributions from each database.}
  \caption{The number of included articles in three databases between 2004-2024.}
  \label{trendsofincludedarticles}
\end{figure}

The first author of the paper (i.e., the coder) manually screened the titles and abstracts of all 2,035 retrieved articles. The coder proceeded to screen articles for inclusion through a three-stage process. First, the coder reviewed the retrieved articles' titles and keywords (i.e., level-one screening). The coder retained articles whose titles or keywords met the inclusion and exclusion eligibility criteria, resulting in 1,365 articles excluded and 670 articles selected for the next stage. Then, in the level-two screening, the coder performed a second review that included the articles' titles, keywords, and abstracts. Based on the inclusion and exclusion criteria, the coder excluded 465 articles and selected 204 articles to read in full. In the third stage (``Eligibility''), the coder read and reviewed each article's main content in full, excluding 142 articles and selecting 62 articles to build the final corpus. This selection process is explained in more detail in the results section and illustrated in Fig.~\ref{prisma}. The coder then presented the selected articles to the other co-authors, who reviewed the coder's classification in a second cycle together. All authors agreed to analyze the 62 articles, and the coder continued with the data extraction and synthesis stage. Fig.~\ref{trendsofincludedarticles} shows the yearly trend of the 62 articles included in our search corpus. The final list of included articles is available in the \textit{Supplementary Materials}.

% ...This left us with 2,035 articles for screening.

\subsection{Data Extraction and Synthesis}
Once the final corpus was established, the coder extracted data to synthesize their characteristics, methods, and findings. Using a Google Spreadsheet, the coder extracted data pertaining to: 
\begin{itemize}
    \item Year of publication
    \item Research type: Bibliographic, Descriptive, Case Study, Laboratory Experiment, Explanatory, or Non-empirical.
    \item Evaluation Methodology: Quantitative, Qualitative, or Mixed Methods.
    \item VR device type if used in the study (i.e., HTC Vive, Oculus Rifts, Meta Quest). 
    % \item Body tracking (i.e., full body or not)
    \item Number of participants recruited in the study. 
    \item Number of participants per group.
    \item Type of experimental task of the study.
\end{itemize}

The coder also used a Google Doc to take open notes and memos that synthesized the key findings of each article. On average, the coder took 15 notes per article. These notes provided more context and details on the technologies, design processes, and results described in each article. These notes are also available in the \textit{Supplementary Materials}.

\subsection{Thematic Analysis}
We conducted a thematic analysis to identify cohesive themes that captured patterns across the final corpus \cite{braun2006using}. The goal of this analysis was to inductively derive themes grounded in the data while minimizing confirmation bias. The coder re-reviewed the articles, including their extracted data and the notes summarizing the main contributions, to generate initial codes. These codes were developed inductively, without reference to pre-existing coding frameworks or theoretical models \cite{Nowell2017,braun2006using}. The codes were not mutually exclusive since the articles could address multiple codes. The coder iterated through the initial codes multiple times, conducting rounds of analysis until no new ideas emerged. Following this, the refined codes were grouped into themes, which were further revised and renamed over several rounds to improve their clarity and alignment with the corpus data. To enhance reliability and interpretive consensus, all the authors reviewed and discussed the themes and their conceptual boundaries in multiple meetings. Ultimately, the final themes were organized into three overarching categories and nine subcategories that captured the breadth and nuance of how Social VR has been studied in the context of collaboration.
\section{Results}
After applying the methodology described above, 62 of the 2,035 papers were included in our final corpus. Fig. ~\ref{prisma} shows the filtering process of the papers in the IEEE Xplore Library, ACM Digital Library, and Web of Science databases. The most common reason for excluding articles during the eligibility stage was that they were not related to collaboration in VR---meaning the study either focused on collaboration without using VR (e.g., in 2D platforms), used VR without addressing collaboration, or addressed neither. This exclusion category is labeled in Fig. ~\ref{prisma} as "not related to VR, collaboration, or both" (\textit{n} = 356). Following this process, Fig.~\ref{trendsofincludedarticles} depicts the trend in the number of papers for each year among the 62 included articles from 2004 to 2024.

\subsection{Description of the Included Articles}
Table \ref{tab:study-characteristics} provides an overview of research and experimental characteristics of the 62 included articles. Table \ref{tab:study-characteristics} summarizes the research types, evaluation methodologies, and analysis methods. The majority were laboratory experiments (69.35\%), followed by non-empirical papers (12.90\%), survey papers (8.07\%), and qualitative descriptive studies (4.84\%). Three articles (4.84\%) fell under other categories, including one case study, one conceptual framework, and one system architecture description.

Regarding evaluation methodology, 33.87\% of the studies employed mixed methods, 32.26\% used quantitative methods, and 17.74\% used qualitative approaches. In 16.13\% of the articles, an evaluation methodology was not employed or specified. Among the qualitative methods articles, thematic analysis was the most common method (19.36\%), followed by free-text responses (4.84\%) and participant experience summaries (3.23\%). A subset of studies employed methods such as open coding, observational techniques, or participant experience summaries. On the quantitative side, ANOVA was the most frequently used analysis (19.28\%), followed by descriptive statistics (15.67\%), \textit{t}-tests and Wilcoxon tests (12.05\% each), and post-hoc or Mann-Whitney U tests (combined 13.24\%). Many studies used more than one quantitative method, so the total count of methods across all papers exceeds 43.

Table \ref{tab:study-characteristics}b summarizes technical and task-related details for the 43 articles that employed laboratory experiments. The most frequently head-mounted display used was the HTC Vive (20.93\%), followed by Oculus Quest 2 (16.29\%), HTC Vive Pro and HTC Vive Pro Eye (9.30\% each), and Oculus Rift (6.99\%). Other headsets, including Meta Quest 3, Oculus Go, and early Oculus Quest models, accounted for smaller proportions.

In terms of types of tasks, problem-solving was the most common task employed in these articles (23.26\%), followed by navigation/ wayfinding (16.28\%) and design/construction (16.28\%). Other tasks included negotiation/planning (11.62\%), meetings/discussions and data analysis (9.30\% each), scientific tasks (6.98\%), and one workplace simulation (2.33\%). Two studies (4.65\%) included task types that did not fall under the primary categories outlined in the table.

Lastly, most articles with laboratory studies examined very small teams (two or three people), which means the findings primarily reflect how small groups work together rather than how larger, more mixed teams might collaborate. Most lab studies also used convenience samples, typically university students. A smaller set recruited professionals (e.g., scientists, designers) for domain-specific tasks, and very few observed real workplace teams doing their everyday work. Reporting on recruitment locale and demographics was often limited. 

\begin{table}[!htb]
  \centering
  \footnotesize
  \setlength{\tabcolsep}{4pt}
  \renewcommand{\arraystretch}{1.0}

  \begin{minipage}[t]{0.48\textwidth}
    \centering
    \begin{tabular}{
      >{\raggedright\arraybackslash}m{3.8cm} 
      >{\raggedleft\arraybackslash}m{1cm} 
      >{\raggedleft\arraybackslash}m{1cm}
    }
      \toprule
      \textbf{Characteristic} & \textbf{Count} & \textbf{(\%)} \\
      \midrule
      \multicolumn{3}{l}{\textbf{Research Type}} \\
      Laboratory Experiment & 43 & 69.35 \\
      Non-empirical & 8 & 12.90 \\
      Survey Paper & 5 & 8.07 \\
      Descriptive (Qualitative) & 3 & 4.84 \\
      Other & 3 & 4.84 \\
      \midrule
      \multicolumn{3}{l}{\textbf{Evaluation Methodology}} \\
      Mixed Methods & 21 & 33.87 \\
      Quantitative & 20 & 32.26 \\
      Qualitative & 11 & 17.74 \\
      N/A & 10 & 16.13 \\
      \midrule
      \multicolumn{3}{l}{\textbf{Qualitative Analysis Method}} \\
      Thematic Analysis & 12 & 19.36 \\
      Free-text Response Analysis & 3 & 4.84 \\
      Participant Experience Summary & 2 & 3.23 \\
      Open Coding Analysis & 1 & 1.61 \\
      Observational Evaluation & 1 & 1.61 \\
      N/A & 43 & 69.35 \\
      \midrule
      \multicolumn{3}{l}{\textbf{Quantitative Analysis Method}} \\
      ANOVA & 16 & 19.28 \\
      Descriptive Statistics & 13 & 15.67 \\
      T-test & 10 & 12.05 \\
      Wilcoxon Test & 10 & 12.05 \\
      Post-hoc Tests & 6 & 7.22 \\
      Mann-Whitney U Test & 5 & 6.02 \\
      Others & 23 & 27.71 \\
      \bottomrule
    \end{tabular}
    \caption*{\centering {a.} Research Types and Analysis Methods (All 62 Studies)}
    \label{tab:study-methods}
  \end{minipage}
  \hfill
  \begin{minipage}[t]{0.48\textwidth}
    \centering
    \begin{tabular}{
      >{\raggedright\arraybackslash}m{3.8cm} 
      >{\raggedleft\arraybackslash}m{1cm} 
      >{\raggedleft\arraybackslash}m{1cm}
    }
      \toprule
      \textbf{Characteristic} & \textbf{Count} & \textbf{(\%)} \\
      \midrule
      \multicolumn{3}{l}{\textbf{VR Device Type}} \\
      HTC Vive & 9 & 20.93 \\
      HTC Vive Pro & 4 & 9.30 \\
      HTC Vive Pro Eye & 4 & 9.30 \\
      Oculus Rift & 3 & 6.99 \\
      Oculus Go & 1 & 2.33 \\
      Oculus Quest & 2 & 4.65 \\
      Oculus Quest 2 & 7 & 16.29 \\
      Meta Quest 3 & 1 & 2.33 \\
      Meta Quest Pro & 1 & 2.33 \\
      Others & 11 & 25.58 \\
      \midrule
      \multicolumn{3}{l}{\textbf{Experimental Task Type}} \\
      Problem Solving & 10 & 23.26 \\
      Navigation/Wayfinding & 7 & 16.28 \\
      Design/Construction & 7 & 16.28 \\
      Data Analysis/Visualization & 4 & 9.30 \\
      Negotiation/Planning & 5 & 11.62 \\
      Meeting/Discussion & 4 & 9.30 \\
      Scientific & 3 & 6.98 \\
      Workplace Simulation & 1 & 2.33 \\
      Other & 2 & 4.65 \\
      \midrule
      \multicolumn{3}{l}{\textbf{Participants per Group}} \\
      Two & 34 & 79.08 \\
      Three & 7 & 16.28 \\
      Four & 1 & 2.32 \\
      Seven & 1 & 2.32 \\
      \bottomrule
    \end{tabular}
    \caption*{\centering {b.} Devices, Tasks, and Group Sizes (43 Lab Studies)}
    \label{tab:study-details}
  \end{minipage}
  
  \Description{Table summarizing characteristics of the 62 included studies. The table lists research types, evaluation methodologies, and qualitative and quantitative analysis methods across all studies, as well as VR device types, experimental task types, and participant group sizes for the subset of laboratory studies.}

  \caption{Overview of research and experimental characteristics of the 62 included articles. Table 1a summarizes the research types, evaluation methodologies, and analysis methods. Table 1b presents VR device types, experimental tasks, and group sizes from 43 lab studies.}
  \label{tab:study-characteristics}
\end{table}

\subsection{Thematic Analysis}
Our thematic analysis revealed three main themes: (1) \textit{``Individual Perception and Experience of Collaboration in Social VR,''} (2) \textit{``Team Dynamics and Collaboration in Virtual Environments,''} and (3) \textit{``Enabling Collaboration through Social VR Affordances''} These themes reflect an increasing focus on employing Social VR to support collaboration by leveraging user experience, team dynamics, and system affordances. Moreover, they uncover significant gaps and future research directions, especially in understanding how specific user interactions and technical elements could effectively enable collaboration in Social VR. 

The first theme includes articles that address the foundational aspects of individual perception and experience within Social VR. This theme includes two sub-themes: \textit{``Am I Really There?,''} which highlights how Social VR can simulate users' physical presence, enhancing their interaction and engagement with the group task. The second sub-theme, \textit{``How Am I Perceived Here?,''} explores avatar design and its impact on collaborative dynamics and engagement, emphasizing the importance of avatar customization to shape user interactions and self-expression within virtual spaces.

The second theme focuses on team dynamics and collaboration in immersive virtual environments, examining how Social VR can enhance team collaboration through social presence. This theme includes four sub-themes: \textit{``Are We Really Together?,''} which explores how social presence contributes to an improved team collaboration; \textit{``Is This Collaboration Real?,''} which addresses how user perceive the authenticity of collaboration in VR; \textit{``Should I Trust this Person?,''} which explores trust-building as a critical element of effective virtual collaboration; and \textit{``Are We Fully Engaged?,''} which considers the role of social engagement in VR collaboration and the different factors that shape user involvement.

The final theme examines how collaboration is supported by the unique affordances of Social VR. These affordances refer to the technical elements and interactions designed to enable realistic and functional 3D virtual environments. This theme includes three sub-themes: \textit{``Creating Realistic and Functional Virtual Environments,''} which addresses how the design of virtual spaces that mimic real-world settings can improve user experience; \textit{``Facilitating Interaction through Virtual Tools and Objects,''} which emphasizes the importance of object manipulation and tool use in collaborative tasks; and \textit{``Supporting Presence through Spatial Audio and Movement,''} which underscores how spatial audio and coordinated movement contribute to a stronger sense of presence and effective collaborative experience. A summary of these themes, sub-themes, and their respective articles can be found in Table~\ref{table:themes}.

\begin{table}[!htb]
\renewcommand{\arraystretch}{1.2}

\footnotesize
\begin{tabular}{p{8.5cm}p{0.7cm}p{3.5cm}}
\toprule
\textbf{Theme} & \textbf{N} & \textbf{Articles} \\
\midrule

\multicolumn{3}{l}{\textbf{Individual Perception and Experience of Collaboration in Social VR.}} \\
``Am I Really There?'' Sense of Presence as a Foundation for Collaboration in Social VR. & 4 & 
\cite{olaosebikan2022identifying}, \cite{khojasteh2021working}, \cite{mei2021cakevr}, \cite{miller2021synchrony} \\
``How Am I Perceived Here?'' Avatar Design Shaping Collaborative Dynamics and Engagement. & 11 &
\cite{osborne2023being}, \cite{torro2023design}, \cite{yang2024exploring}, \cite{yassien2021give}, \cite{kimmel2024kinetic}, \cite{kocur2020effects}, \cite{yassien2020design}, \cite{kallioniemi2017effect}, \cite{shoa2023sushi}, \cite{li2024social}, \cite{young2015dyadic} \\[0.5cm] \hline

\multicolumn{3}{l}{\textbf{Team Dynamics and Collaboration in Virtual Environments.}} \\
``Are We Really Together?'' Social Presence Enhancing Team Collaboration. & 12 &
\cite{olaosebikan2022identifying}, \cite{dey2024social}, \cite{darbar2024gazemolvr}, \cite{bovo2022cone}, \cite{ide2020effects}, \cite{adkins2024hands}, \cite{mei2021cakevr}, \cite{kimmel2023let}, \cite{wei2022communication}, \cite{weissker2023enhanced}, \cite{moharana2022subjective},\cite{chastine2005ammp} \\
``Is This Collaboration Real?'' Perceiving Realness in Virtual Teamwork. & 6 &
\cite{torro2023design}, \cite{maloney2020falling}, \cite{osborne2023being}, \cite{bailenson2006longitudinal}, \cite{park2024lessons}, \cite{dey2024social} \\
``Should I Trust this Person?'' Enabling Trust as the Glue that Binds Collaborators. & 2 &
\cite{torro2023design}, \cite{yassien2020design} \\
``Are We Fully Engaged?'' The Importance of Social Engagement in VR Collaboration. & 7 &
\cite{xu2023designing}, \cite{le2024impact}, \cite{moharana2023physiological}, \cite{moore2019communicating}, \cite{gumilar2022inter}, \cite{duchowski2004visual}, \cite{yang2022towards} \\[0.5cm] \hline 

\multicolumn{3}{l}{\textbf{Enabling Collaboration through Social VR Affordances.}} \\
%\multicolumn{3}{l}{\textbf{Technical Elements and User Interactions.}} \\[0.5em]
Creating Realistic and Functional Virtual Environments. & 9 &
\cite{mei2021cakevr}, \cite{osborne2023being}, \cite{men2019lemo}, \cite{freiwald2020conveying}, \cite{weissker2021group}, \cite{langa2022multiparty}, \cite{freeman2022working}, \cite{saffo2021remote}, \cite{zhang2023vrgit} \\
Facilitating Interaction through Virtual Tools and Objects. & 11 &
\cite{garcia2008enhancing}, \cite{garcia2019collaborative}, \cite{dey2024social}, \cite{osborne2023being}, \cite{roberts2005supporting}, \cite{roberts2004supporting}, \cite{dos2011collaborative}, \cite{otto2006review}, \cite{nguyen2017collavr}, \cite{argelaguet2011see}, \cite{rasch2024just} \\
Supporting Presence through Spatial Audio and Movement. & 10 &
\cite{weissker2019multi}, \cite{immohr2024evaluating}, \cite{immohr2024subjective}, \cite{rios2018users}, \cite{immohr2023proof}, \cite{tim2020getting}, \cite{naef2004blue}, \cite{deacon2023invoke}, \cite{rasch2023going}, \cite{sarmiento2014measuring} \\

\bottomrule
\end{tabular}
\Description{Table listing themes and sub-themes identified through thematic analysis of Social Virtual Reality collaboration research. Each row shows a sub-theme, the number of associated articles, and corresponding references, grouped under three main theme categories: individual perception and experience, team dynamics and collaboration, and collaboration enabled through Social VR affordances.}
\caption{Themes and sub-themes identified through thematic analysis of Social VR research on collaboration. Each row presents a sub-theme, the number of articles (N) associated with it, and its references. A total of 72 entries are shown, as some papers contribute to multiple themes.}

\label{table:themes}
\end{table}

\subsubsection{Individual Perception and Experience of Collaboration in Social VR}
\paragraph{``Am I Really There?'' Sense of Presence as a Foundation for Collaboration in Social VR}
\label{subtheme_am_I_really_there}
This sub-theme focuses on individual users' sense of presence---the subjective feeling of ``being there'' within a 3D virtual environment \cite{triberti2025being}---as a foundational condition for collaboration in Social VR. Distinct from social presence (i.e., perceiving others), \textit{presence} describes the user’s own experience of being located in the virtual space. This experience is shaped by how intuitively users' actions and intentions are translated into the environment \cite{oh2018systematic,olaosebikan2022identifying,miller2021synchrony}.

Studies reported mixed effects of presence: while some found VR increased presence and task engagement compared to video calls \cite{olaosebikan2022identifying,mei2021cakevr}, others observed flat presence scores despite growing familiarity \cite{khojasteh2021working}. These inconsistencies suggest that presence depends less on time in VR and more on the availability of natural nonverbal cues. This helps explain the lack of increase: adaptation alone did not lift presence because natural nonverbal channels were missing or awkward. Consistently, Mei et al. \cite{mei2021cakevr} propose adding emoji-like or haptic cues to convey emotion, and report that missing facial expressions weakened users’ emotional connection. Across these papers, gaps in nonverbal expression and practical hurdles (learning curve, limited in-VR tools) made it harder to feel ``there'' during collaboration \cite{olaosebikan2022identifying,khojasteh2021working,mei2021cakevr}.

% Across studies, results are mixed: scientists collaborating in VR reported higher presence than in video calls and valued arranging information on room surfaces and grabbing/rotating data objects \cite{olaosebikan2022identifying}; similarly, in a co-design tool, participants felt “in the task” when they could move and shape shared 3D elements \cite{mei2021cakevr}. However, in a five-session study, people became more comfortable with VR, yet presence scores stayed flat \cite{khojasteh2021working}. The authors note that participants relied mainly on voice and word choice to read emotion, and a button-based avatar emotion feature felt unnatural and was used little; they point to real-time face tracking as a more promising direction \cite{khojasteh2021working}. 

%In sum, individual presence supports collaboration when systems allow intuitive action and natural expression; without these, getting comfortable with VR alone does not raise presence.

\paragraph{``How Am I Perceived Here?'' Avatar Design Shaping Collaborative Dynamics and Engagement.}
\label{subtheme_how_am_I_perceived}
This sub-theme examines how avatar design influences users' experiences in Social VR. We found that multiple articles focused on both how users perceive their own virtual bodies (i.e., self-perception) and how they believe others perceive them. Their main findings indicate that avatar embodiment plays a key role in how users express themselves, interpret social cues, and engage in collaborative settings. The visual and behavioral fidelity of avatars impacts users' sense of identity and presence, influencing their confidence, expressiveness, and overall engagement during collaboration.

The articles of this corpus have demonstrated that avatar characteristics---such as gender, realism, abstraction level, and humanoid forms---significantly shape users' experiences \cite{osborne2023being,shoa2023sushi,torro2023design,yang2024exploring,kallioniemi2017effect,li2024social,young2015dyadic}. Gender representation emerged as a recurrent aspect across these studies. For instance, Yassien et al. \cite{yassien2021give} found that having all the group members' avatars with the same gender (e.g., all-male or all-female) improved collaboration quality and mutual support, even when users' real-world genders differed. Similarly, Yang et al. \cite{yang2024exploring} demonstrated that users represented by non-human avatars (e.g., fox, robot, duck) reported greater comfort and equal participation than when using communication formats where their real-world appearance was visible, such as face-to-face or video conferencing settings, suggesting that abstract representations may alleviate social pressure and enable more balanced collaboration.

Two studies emphasized the importance of avatar customization and stylization when collaborating in Social VR \cite{osborne2023being,freeman2020my}. Osborne et al. \cite{osborne2023being} showed that allowing users to personalize their avatars enhanced engagement with the rest of the group. Similarly, Freeman et al. \cite{freeman2020my} demonstrated that enabling self-representation in Social VR supports identity exploration, particularly for users navigating gender or sexual identity. Participants often described their avatars as extensions of themselves, using them to explore self-image in ways that felt safe and empowering \cite{freeman2020my}.

However, some articles showed that increased realism in avatar design can complicate users' sense of self-identification by creating a mismatch between the avatar’s appearance and the user’s internal self-image. Two studies addressed how the ``uncanny valley'' effect---which occurs when avatars look nearly human but still exhibit subtle imperfections in appearance or movement---made participants feel unnatural or unsettled during the collaboration experience \cite{osborne2023being,torro2023design}. Osborne et al. \cite{osborne2023being} and Torro et al. \cite{torro2023design} reported that these unsettling avatars disrupted users' sense of presence and hindered their natural interactions. 

Several articles examined how the fidelity of avatars' motion and expression influences users' perceptions of being accurately represented in their social interactions. Kimmel et al. \cite{kimmel2024kinetic} and Yassien et al. \cite{yassien2020design} found that full-body avatars with realistic tracking and expressive features helped users feel more accurately represented. Similarly, Kocur et al. \cite{kocur2020effects} found that mismatches between how individuals wanted to present themselves and how they believed others perceived them---often due to avatar limitations---undermined users' confidence and reduced their engagement in group collaboration.

In summary, these articles demonstrate that avatars are not merely cosmetic features but critical components of users' collaborative experiences in VR. They influence how individuals perceive themselves, their social roles, expressions, and feelings of being acknowledged within shared spaces. 

\subsubsection{Team Dynamics and Collaboration in Virtual Environments}

\paragraph{``Are We Really Together?'' Social Presence Enhancing Team Collaboration.}
\label{subtheme_are_we_really_together}
This sub-theme explores how \textit{social presence}---the sense of ``being with others'' in a shared virtual environment \cite{oh2018systematic}---facilitates effective team collaboration in Social VR. While the sense of presence refers to a user's perception of presence, social presence refers to the awareness of and connection to teammates. It addresses the feeling that others are truly co-located and responsive within the virtual space.

Several articles emphasized that social presence is a foundational social-cognitive state that enables key collaborative behaviors such as turn-taking, mutual coordination, and shared understanding \cite{olaosebikan2022identifying,moharana2022subjective,dey2024social,kimmel2023let,wei2022communication,weissker2023enhanced,chastine2005ammp}.  A key component of social presence in VR is mutual attention, which is the ability to perceive others' attention and actions in real-time \cite{darbar2024gazemolvr,bovo2022cone,ide2020effects,weissker2023enhanced}. For example, Darbar et al. \cite{darbar2024gazemolvr} introduced `GazeMolVR,' a system for molecular visualization in collaborative VR that incorporated shared eye-gaze cues. Their findings showed that real-time gaze sharing improved the feeling of connectedness and eased conversational flow. When collaborators saw others’ gaze, they adjusted their own attention more fluidly, making the interaction feel more natural and co-present \cite{darbar2024gazemolvr}.

The articles also demonstrate that social presence is strongly connected to nonverbal awareness and emotional connection. When avatars or systems convey cues such as gaze direction, pointing, or simple gestures, people can more easily interpret each other's intentions \cite{bovo2022cone,ide2020effects}. For example, studies such as Bovo et al. \cite{bovo2022cone} and Ide et al. \cite{ide2020effects} implemented visual aids, including cones of vision and symbolic gestures, to help dyads perceive their collaborators' focus and actions. These features not only added expressiveness but also enhanced social presence by increasing mutual visibility and acknowledgment, helping users feel noticed and understood by others.

One article highlighted that social presence enhances coordination during multi-party interactions \cite{adkins2024hands}. In their evaluation of audio and interaction modalities, Adkins et al. \cite{adkins2024hands} found that dyads using spatial audio, compared to non-spatial conditions, experienced smoother turn-taking and reported feeling more ``in sync'' with one another. These findings suggest that social presence is shaped not only by visual or avatar-based cues but also by auditory and behavioral alignment that supports dyadic coordination.

One last article noted that weak or underdeveloped social presence negatively affects collaboration. In their evaluation of a co-design task, Mei et al. \cite{mei2021cakevr} found that the absence of avatar reactivity and expressive behavior made interactions feel more like a phone call rather than a shared, embodied collaboration. This diminished sense of social presence reduced user engagement and made communication feel less natural, highlighting the importance of social presence in enabling teamwork.

In summary, these articles demonstrate that social presence serves as the social glue that makes collaboration in Social VR feel natural and connected. Ultimately, social presence enables members to perceive and respond to one another's actions in real time, supporting cohesive and authentically shared collaboration in VR. 
 
\paragraph{``Is This Collaboration Real?'' Perceiving Teamwork' Realness in Social VR}
\label{subtheme_is_this_collaboration_real}
This sub-theme includes articles that investigate how users evaluate the authenticity, seriousness, and naturalness of their collaborative experiences in Social VR. The focus is on whether immersive VR environments support interactions that are not only technically possible but also feel socially valid, natural, and meaningful. Beyond the technical functionality, these articles highlight that users carry expectations shaped by physical-world norms, emotional cues, and social dynamics, which influence how ``real'' collaboration feels in Social VR \cite{torro2023design,maloney2020falling,osborne2023being}. 

A key factor in how users evaluate the authenticity of virtual collaboration is whether their partners appear as genuine social actors \cite{torro2023design}. Torro et al. \cite{torro2023design} introduced the concept of the ``social presence illusion'' to describe how lifelike and responsive avatars make interactions feel more socially authentic. VR systems that support behavioral realism---such as timely gaze, natural gestures, and coordinated responses---enable familiar social rituals (e.g., turn-taking, nonverbal responses) that help elevate collaboration from a purely functional exchange into an interaction that feels intuitively real \cite{torro2023design,bailenson2006longitudinal}.

Building on this, Maloney et al. \cite{maloney2020falling} interviewed long-term Social VR users to understand what made virtual interactions feel meaningful. Participants reported that their collaborations felt more authentic when gestures, feedback, and shared activities aligned with everyday social norms. Many of them expressed frustration with unrealistic avatar features---such as exaggerated cartoon-like hands or faces that could not convey realistic emotions---which made the interactions feel artificial and less socially authentic. What users desired was not just visual realism, but interaction fidelity, which translated into responsiveness and expressiveness that matched the tone and rhythm of real-world collaboration.

Other articles in this sub-theme indicated that misaligned avatar behavior can disrupt the authenticity of collaboration. For example, Osborne et al. \cite{osborne2023being} noted that technical glitches, such as avatar duplication or unnatural expressions, made interactions feel artificial and awkward. In contrast, when avatars moved and responded in socially expected ways, participants perceived the collaboration as smooth and believable, even when the avatars were abstract in design \cite{osborne2023being}.

Beyond appearance, one article explored how subtle behavioral manipulations could enhance the cohesion and authenticity of collaboration in Social VR \cite{bailenson2006longitudinal}. In their foundational work, Bailenson et al. \cite{bailenson2006longitudinal} introduced the concept of ``Transformed Social Interaction'' (TSI), demonstrating that altering avatar behaviors (e.g., gaze direction, synchronized gestures) can increase perceived team unity. Their findings suggest that collaborative realism depends not only on shared goals or tasks established in the virtual environment but also on how interaction feels on a social level. By aligning nonverbal cues with social expectations, Social VR applications can strengthen the psychological sense of working as a single team \cite{bailenson2006longitudinal}.

One study examined how to reinforce the role of realism in a Social VR workplace by eliminating fantastical elements.  Park et al. \cite{park2024lessons} removed the teleportation feature and required users to navigate through halls and elevators, mirroring how people would walk in a physical office. While this approach increased social encounters and enhanced users' social presence, it also introduced usability challenges, revealing a tension between realism and practicality \cite{park2024lessons}. The findings of this article echo broader concerns about the necessary levels of realism to foster authentic collaboration without compromising the unique affordances of VR.

Lastly, one article \cite{dey2024social} demonstrated that Social VR participants who first engaged in an emotionally intense solo experience approached a subsequent group task with more seriousness and coordination. The realism of the initial simulation made the following collaboration feel more consequential and authentic. Dey et al.'s findings \cite{dey2024social} suggest that initial VR experiences can reshape how users' perceptions of group work in Social VR, making virtual collaboration feel more meaningful, important, and socially grounded.

In summary, these articles demonstrate that when Social VR users feel that their actions have an impact and that their collaborators are genuinely engaged, the experience transcends the simulation and becomes a psychologically meaningful, socially credible form of collaboration.

\paragraph{``Should I Trust this Person?'' Enabling Trust in Social VR}
\label{subtheme_should_i_trust}
This sub-theme examines how trust, an essential yet often overlooked foundation of effective collaboration, is developed in Social VR. While trust in co-located settings often emerges through informal interactions and subtle nonverbal cues, trust-building mechanisms are often limited or absent in remote communication. The reviewed articles show that Social VR can restore some of these cues through embodied interaction, avatar behavior, and shared spatial context \cite{torro2023design,yassien2020design}.

Torro et al. \cite{torro2023design} proposed a design theory explaining how Social VR can support trust-building in virtual teams. Grounded in social exchange theory---which emphasizes reciprocal interactions as the foundation of relationships \cite{cook2013social}---they proposed that trust emerges through two key pathways in Social VR: a \textit{peripheral} route, driven by first impressions (e.g., avatar appearance, social cues), and a \textit{central} route, based on deeper assessments of a partner's competence, integrity, and goodwill \cite{torro2023design}. For example, avatars that mimic natural gestures, proximity-based conversations, and realistic movement can help promote social bonding. Moreover, features like virtual break rooms with spatial properties can support casual, unstructured interactions, which are often absent in structured video calls or text-based platforms.

Yassien et al. \cite{yassien2020design} highlighted another key dimension of trust in Social VR: personal space. This article found that users felt safer and more connected when their virtual personal space was respected. Trust was undermined when avatars came too close or moved in ways that felt erratic or unpredictable, causing discomfort and breaking the sense of social ease. The researchers recommended designing Social VR applications with mechanisms to prevent collisions and unwanted proximity between avatars, reinforcing the importance of comfort and perceived safety.

Together, these two articles suggest that trust in Social VR is supported through both social signaling and environmental design, including expressive avatars, respectful personal space, and opportunities for informal interaction. 

\paragraph{``Are We Fully Engaged?'' The Importance of Social Engagement in VR Collaboration.}
\label{subtheme_are_we_fully_engaged}
This sub-theme focuses on social engagement, the extent to which collaborators in Social VR remain emotionally and communicatively connected during a shared task \cite{xu2023designing,yang2022towards}. Social engagement goes beyond mere attention or performance; it reflects how present, expressive, and emotionally involved individuals feel in relation to their teammates \cite{xu2023designing,yang2022towards}. The reviewed articles identified design strategies that support behavioral cues of engagement and help sustain interpersonal connections throughout the Social VR collaboration \cite{le2024impact,xu2023designing,moore2019communicating,moharana2023physiological,gumilar2022inter,duchowski2004visual}.

One strategy identified in this sub-theme involves enhancing avatar behaviors to foster social engagement. Le et al. \cite{le2024impact} introduced an augmented avatar system that generated subtle nonverbal cues---such as head nods and posture shifts---even when users did not perform them directly. These enhancements made collaborators appear more responsive and involved \cite{le2024impact}. Similarly, Xu et al. \cite{xu2023designing} studied older adults co-planning virtual travel experiences using VR and found that social engagement deepened when participants shared personal memories and life stories related to the destinations. These moments of self-disclosure fostered warmth and mutual interest, highlighting that social engagement can be supported not only through interface design but also through content that invites personal connection and emotional expression.

Another article explored physiological synchrony as a deeper measure of engagement. Moharana et al. \cite{moharana2023physiological} tracked heart rate patterns during a collaborative memory game and found that dyads with higher physiological synchrony reported greater feelings of connection, presence, and workflow. Their findings suggest that social engagement may operate at both cognitive and emotional levels and that VR environments can help capture or even facilitate these more subtle social dynamics.

Taken together, these articles show that collaboration in Social VR extends beyond visual realism or task efficiency. It hinges on participants feeling seen, acknowledged, and emotionally connected, which are qualities that can be nurtured through expressive avatars, thoughtful interaction design, and moments that invite genuine interpersonal connections. 

\subsubsection{Enabling Collaboration through Social VR Affordances}
\paragraph{Creating Realistic and Functional Virtual Environments.}
\label{subtheme_creating_realistic}
This sub-theme explores how the spatial layout, visual realism, and responsiveness of the virtual environment impact users' interactions and support collaboration in Social VR. Rather than focusing on tools or input mechanisms, the reviewed articles emphasize how the environment's visual design, structure, and adaptability influence where and how users interact. The findings suggest that effective virtual environments in Social VR balance familiar spatial cues with flexible design elements, enabling collaborative behaviors that can surpass those possible in the physical world \cite{mei2021cakevr,osborne2023being,men2019lemo,freiwald2020conveying,weissker2021group,langa2022multiparty,freeman2022working,saffo2021remote,zhang2023vrgit}.

Two articles found that environmental realism is most valued when it reinforces familiar spatial dynamics \cite{mei2021cakevr,osborne2023being}. Mei et al. \cite{mei2021cakevr} found that the absence of environmental sound, realistic tools, or background detail made the virtual workspace feel less functional to users. However, the articles discussed that greater realism does not necessarily lead to a better user experience. Osborne et al. \cite{osborne2023being} proposed a typology of Social VR environments to support users' engagement: \textit{Skeuomorphic}, environments designed as realistic office-like spaces; \textit{Experimental}, environments designed as creative and open-ended spaces, and \textit{Prefab-based}, which are template-driven but customizable by the users. Their findings suggest that adaptability is often more effective than fidelity. Therefore, applications in Social VR should be tailored to users' goals and team dynamics to better support their collaborations, instead of focusing solely on visual realism.

Research explains how layout and spatial configuration shape users' access, privacy, and coordination in Social VR \cite{freeman2022working,saffo2021remote,men2019lemo,zhang2023vrgit}. In a study on collaborative music creation, Men et al. \cite{men2019lemo} found that participants favored publicly visible personal spaces that balanced individual work zones with group awareness. They also revealed trade-offs in spatial arrangements: positioning avatars side-by-side offered shared perspectives but limited equal access to shared interactive virtual tools, while face-to-face setups improved visibility for all participants but required redesigned interfaces to maintain usability. In sum, spatial affordances---such as visibility, proximity, and orientation---can directly impact collaborative dynamics in Social VR applications.

Two articles highlighted how Social VR's affordances---such as spatial layout and 3D visibility cues---support team coordination by aligning users' viewpoints and actions \cite{freiwald2020conveying}. Freiwald et al. \cite{freiwald2020conveying} compared three visual indicators to represent users' viewpoints during a collaborative task. They found that displaying a live feed of each user's view reduced their task's error rates, while adding a 3D cone to indicate the boundaries of each user's viewpoint enhanced their social presence. Similarly, Weissker et al. \cite{weissker2021group} developed a group navigation technique that allowed collocated users to teleport together while preserving their relative positions. This feature helped teams stay spatially coordinated as they explored larger virtual environments \cite{weissker2021group}.

Lastly, one study addressed the technical setup required to scale larger groups and collaborative spaces in Social VR. Langa et al. \cite{langa2022multiparty} developed a VR application for multiparty volumetric meetings, focusing on accessibility and stability for larger groups. While their evaluation involved groups of two and four, the system was designed to scale beyond small teams without sacrificing responsiveness or presence. Their use of volumetric user representations---captured and rendered in 3D---enabled more lifelike, embodied interactions than traditional avatars, offering a medium for realistic and scalable Social VR collaboration \cite{langa2022multiparty}.

In summary, these articles suggest that the most effective Social VR applications do not simply mimic the real world. Instead, they introduce new affordances that are not possible to afford in face-to-face meetings---such as view-sharing, spatial partitioning, and adaptable layouts---to support effective collaboration. 

\paragraph{Facilitating Interaction through Virtual Tools and Objects.}
\label{subtheme_facilitating_interaction}
Articles in this sub-theme examine how virtual tools and manipulable objects enable collaboration in Social VR. Unlike traditional online collaboration platforms that rely on screen-based interactions, Social VR allows users to engage with shared tools using hand movements, gaze direction, and spatial movement, making teamwork feel more tangible and embodied. Across the reviewed articles, tools such as pens, whiteboards, 3D models, and laser pointers were not only instrumental to complete tasks but also served a social function, enabling users to coordinate actions, maintain mutual awareness, and collaboratively build shared outcomes \cite{garcia2008enhancing,garcia2019collaborative,dey2024social,osborne2023being,dos2011collaborative,nguyen2017collavr,rasch2024just}.

Previous research has closely examined coupled collaborative tasks that required users to coordinate in real-time by synchronizing and sequentially manipulating shared objects \cite{otto2006review,roberts2005supporting,roberts2004supporting,garcia2008enhancing}. Roberts et al. \cite{roberts2005supporting} designed a Gazebo-building task in which participants had to pass, plan, and co-assemble components in real-time. Garcia et al. \cite{garcia2008enhancing} addressed the technical challenges of network consistency and inconsistent feedback synchronization across users in VR by introducing visual cues---such as arrows and fall indicators---that helped participants anticipate object behavior and reduce errors during their interactions.

Other articles emphasized the importance of tool accessibility and ease of use for effective collaboration in Social VR \cite{rasch2024just,osborne2023being,dos2011collaborative}. While certain features in VR (e.g., sticky notes, drawing tools) supported brainstorming and shared design tasks, platform-specific constraints (e.g., restricted permissions, complex setup processes) often hindered the collaboration experience \cite{osborne2023being,dos2011collaborative,nguyen2017collavr,argelaguet2011see}. Osborne et al. \cite{osborne2023being} noted that AltspaceVR provided relatively few built-in tools for groups and frequently relied on external software, which sometimes disrupted users' interaction flow. Complementing these results, Dey et al. \cite{dey2024social} underscored how direct interaction with virtual objects in Social VR enhances the sense of presence, making the workspace feel more tangible. Their findings suggest that physical engagement with virtual tools not only aids task performance but also strengthens users' sense of connection to the team and the task \cite{dey2024social}.

Taken together, these articles demonstrate that the design and usability of virtual tools directly affect the collaborative dynamics in Social VR. Intuitive and accessible tools can support effective teamwork by facilitating smoother interactions, idea exchange, and action coordination in coupled tasks. Applications that enable multiple users to jointly control, annotate, or manipulate virtual objects promote more balanced participation and facilitate fluid exchange of ideas. 

\paragraph{Supporting Presence through Spatial Audio and Movement.}
\label{subtheme_supporting_presence}
The last sub-theme examines how audio and movement features in Social VR contribute to a deeper sense of presence and facilitate user coordination. Affordances such as spatial audio and group locomotion help users feel oriented, responsive, and connected within the shared virtual environment, supporting both awareness and collaboration \cite{weissker2019multi,immohr2024evaluating,immohr2024subjective,rios2018users}.

The identified articles demonstrated that spatial audio helps anchor users in the virtual environment by simulating the direction and proximity of voices and ambient sounds \cite{deacon2023invoke,weissker2019multi}. Weissker et al. \cite{weissker2019multi} found that using binaural spatial audio---which simulates 3D sound perception by delivering slightly different signals to each ear---improved triadic collaboration by enhancing social presence, turn-taking behavior, and attentional alignment. Participants reported being more aware of their partner’s position and smoother conversational flow compared to non-spatial audio conditions \cite{weissker2019multi}. However, Immohr et al. \cite{immohr2023proof} found that spatial audio did not significantly improve performance in a negotiation task, suggesting that the effectiveness of audio design depends on the type of task and the dynamics of user interaction. The processing and presentation of spatial audio also influence users’ sense of presence and the credibility of virtual interactions. Immohr et al. \cite{immohr2024evaluating} evaluated different binaural rendering methods and found that more realistic spatial cues improved both the perceived authenticity of communication and the naturalness of user responses. These subtle auditory refinements encouraged more lifelike exchanges, particularly in tasks where directional cues and conversational nuance were critical.

Movement also plays an important role in collaborative tasks within Social VR, particularly those involving co-navigation or shared spatial exploration. Rios et al. \cite{rios2018users} found that synchronizing avatar footstep sounds and animations with users' real movements enhanced spatial awareness. Participants navigated more accurately and walked closer to objects and collaborators. Although participants' subjective sense of presence did not change significantly, behavioral indicators suggested stronger alignment between users and their avatars \cite{rios2018users}.

The reviewed articles introduced a range of group navigation techniques to address the challenges of coordinating movement in collaborative VR applications \cite{tim2020getting,naef2004blue,weissker2019multi,rasch2023going,sarmiento2014measuring}. For instance, Weissker et al. \cite{tim2020getting} developed a group teleportation system that preserved user formations during their jumps. Building on this idea, Naef et al. \cite{naef2004blue} developed the ``blue-c API,'' which enabled synchronized group movement, allowing users to navigate the virtual environment together without losing cohesion. Weissker et al. \cite{weissker2019multi} developed ``Multi-Ray Jumping,'' a system designed for collocated groups that offered directional previews, seamless teleportation, and enhanced mutual awareness during shared navigation. These articles show that group movement in VR is not only about getting from one place to another but also about maintaining spatial coordination and shared context to support effective teamwork \cite{naef2004blue,weissker2019multi}.

In summary, these articles illustrate how spatial audio and group movement features enhance the authenticity and fluidity of collaboration in Social VR. When thoughtfully designed, these elements can improve how teams communicate, stay together, and navigate virtual spaces using Social VR.
\section{Discussion}
This review synthesizes two decades of work on Social VR for collaboration. We identified three overarching conceptual levels that capture how prior work conceptualized collaboration in Social VR (see Table~\ref{description}): \textit{individual experience} (i.e., presence, identity/avatars, comfort), \textit{team dynamics} (i.e., social presence, collaboration realness, trust, engagement), and \textit{affordances} (i.e., space/layout, shared tools/objects, audio and movement). Although patterns emerged, the findings were mixed: some studies reported stronger feelings of ``being there'' and smoother coordination, while others noted flat presence scores, uncanny or awkward avatar behavior, and task-dependent effects of audio and movement. These results underscore the need to examine when Social VR helps or falls short.

In the following subsections, we discuss the key findings of this review with two goals: to identify the intersections and open debates that emerge from these identified studies, and to critically discuss the potential benefits and limitations of leveraging Social VR to support effective collaboration. 

\begin{table}[!htb]
\centering
\footnotesize
\begin{tabular}{@{}>{\raggedright}p{0.19\textwidth}>{\raggedright}p{0.23\textwidth}p{0.54\textwidth}@{}}
\toprule
\makecell[l]{\textbf{Themes}} & \textbf{Description} & \textbf{Key Insights} \\ \hline
\midrule

\multicolumn{3}{@{}p{\textwidth}@{}}{\textbf{Individual Perception and Experience}} \\
[0.3em]

\textit{Am I Really There?} &
  Examines users' sense of presence as a basis for collaboration &
  \vspace{-1em}
  \begin{itemize}[leftmargin=7pt,noitemsep,topsep=-10pt] 
        \item Presence improves with intuitive action and natural expression.
        \item VR adaptation alone did not raise presence in some studies.
        \item Missing nonverbal cues and tool friction reduce feeling 'there'.
    \end{itemize} 
  \\

\textit{How Am I Perceived Here?} &
  Explores how avatar design shapes collaboration &
  \vspace{-1em}
  \begin{itemize}[leftmargin=7pt,noitemsep,topsep=-10pt] 
        \item Customization supports engagement and identity exploration.
        \item Near-human realism can trigger discomfort (uncanny valley).
        \item Mismatch between self-presentation and perceived image lowers confidence.
    \end{itemize} \\ \hline

\multicolumn{3}{@{}p{\textwidth}@{}}{\textbf{Team Dynamics and Collaboration}} \\
[0.3em]

\textit{Are We Really Together?} &
  How social presence enhances teamwork &
  \vspace{-1em}
  \begin{itemize}[leftmargin=7pt,noitemsep,topsep=-10pt] 
        \item Shared gaze/gestures/spatial audio aid turn-taking and mutual awareness.
        \item Emotional connection rises with visible, timely nonverbal cues.
        \item Weak social presence makes teamwork feel like a phone call.
    \end{itemize} \\

\textit{Is This Collaboration Real?} &
  Perception of authenticity in virtual teamwork &
  \vspace{-1em}
  \begin{itemize}[leftmargin=7pt,noitemsep,topsep=-10pt] 
        \item Behavioral realism (timely gaze, responsive motion) matters more than looks.
        \item Transformed Social Interaction can increase cohesion when cues are aligned.
        \item Realistic replication can boost authenticity but also introduce usability challenges.
    \end{itemize} \\

\textit{Should I Trust This Person?} &
  Trust-building in Social VR &
  \vspace{-1em}
  \begin{itemize}[leftmargin=7pt,noitemsep,topsep=-10pt] 
        \item Trust has received little attention in Social VR research, with only two direct studies to date.
        \item Trust built through social cues and personal space.
        \item Informal interactions (e.g., break areas) help bonding.
    \end{itemize}\\

\textit{Are We Fully Engaged?} &
  Social engagement during collaboration &
  \vspace{-1em}
  \begin{itemize}[leftmargin=7pt,noitemsep,topsep=-10pt] 
        \item Augmented avatar cues can raise perceived engagement.
        \item Sharing personal content or self-disclosure deepens connection.
        \item Physiological synchrony (e.g., heart rate) shows deeper engagement.
    \end{itemize}\\ \hline

\multicolumn{3}{@{}p{\textwidth}@{}}{\textbf{Enabling Collaboration through Social VR Affordances.}} \\
[0.3em]

\textit{Creating Realistic and Functional Virtual Environments} &
  Design of spaces for work together &
  \vspace{-1em}
  \begin{itemize}[leftmargin=7pt,noitemsep,topsep=-10pt] 
        \item Adaptability matters more than realism; go beyond replication.
        \item View-sharing and room layouts help balance private work with group coordination.
        \item Customization enhances collaboration effectiveness.
    \end{itemize} \\
    
\textit{Facilitating Interaction through Virtual Tools and Objects.} &
  Object manipulation and tool use &
  \vspace{-1em}
  \begin{itemize}[leftmargin=7pt,noitemsep,topsep=-10pt] 
        \item Tools support both task work and coordination.
        \item Multi-user authoring, annotation, and export/persistence matter.
        \item Platform constraints and setup complexity break flow.
    \end{itemize}\\

\textit{Supporting Presence through Spatial Audio and Movement} &
  Role of spatial audio and locomotion &
  \vspace{-1em}
  \begin{itemize}[leftmargin=7pt,noitemsep,topsep=-10pt] 
        \item Spatial audio improves awareness and turn-taking, but task effects vary.
        \item Group teleportation keeps teammates aligned.
        \item Movement cues can align behavior even when presence scores stay flat.
    \end{itemize}\\

\bottomrule
\end{tabular}
\Description{Table summarizing three overarching themes and associated sub-themes identified in the literature on Social Virtual Reality collaboration. For each sub-theme, the table lists a brief description of the focus area and a set of key insights reported across prior studies, organized under the categories of individual experience, team dynamics, and collaboration affordances.}
\caption{Overview of Themes and Sub-themes with Descriptions and Key Insights.}
\label{description}
\end{table}

\subsection{RQ1: Diverse Research from Design to Evaluation in Social VR}
Regarding how previous studies have addressed team collaboration in Social VR (RQ1), our corpus reveals the potential of Social VR as a distinct medium enabling embodied group interactions unfolding online. These articles diverge from other major strands of VR research, which focus on gaming or education, and emphasize the importance of fidelity for collaboration at the individual, team, and environmental level. The surge in research on collaborative Social VR since 2022--- the COVID-19 pandemic and the renewed interest in `metaverse' technologies---has spurred both novel systems and theoretical conversations about teamwork in immersive 3D environments. 

Our review highlights how core constructs in VR---such as presence, social presence, and the uncanny valley---shape collaboration in Social VR. For example, while the uncanny valley is a known VR challenge, the reviewed articles show how subtle imperfections in avatar realism can disrupt trust, make users feel socially uneasy, and reduce team cohesion, less evident in entertainment contexts. While these concepts were traditionally used in HCI and communication studies to assess user experience, their use is expanding to multi-user contexts. Bridging these constructs with perspectives from team science \cite{mathieu2017century} and groups in CSCW \cite{harris2019} can offer new ways to understand social dynamics in immersive environments.

When assessing what supports collaboration in Social VR, these studies converge on a mechanism that we term \emph{interaction fidelity}. Unlike gaming or training, which can afford lower fidelity, collaboration in Social VR depends on higher realism in avatars and environments, approaching in-person expectations \cite{ball2022metaverse}. The teams from these studies benefited when Social VR allowed them to arrange information in space, share viewpoints, and manipulate common objects with predictable, low-latency responses. They struggled when hand alignment, locomotion, or tools were inconsistent. Avatar design showed a similar trend: expressive and legible behaviors enabled natural interaction, whereas near-human appearances without matching behavior trigger discomfort. Affordances were also critical. Audio and movement features were most useful when they clarified members' attention and turn-taking, though effects were task-dependent. For example, spatial audio improved coordination in group discussions, yet offered little benefit in negotiation tasks. Overall, these results suggest that Social VR is more effective for tasks that hinge on spatial arrangement and embodied interaction---such as manipulating data objects \cite{olaosebikan2022identifying} or co-designing artifacts \cite{mei2021cakevr}---while providing fewer advantages for tasks where such affordances matter less, such as negotiation or conversation tasks \cite{immohr2023proof}.

\subsection{RQ2: Gaps and Future Directions}
The corpus also reveals theoretical and methodological gaps in Social VR research (RQ2). First, most empirical studies involved dyads and small groups in short periods. While these studies have shed light on presence, trust, and coordination, they raise questions about scalability and long-term effects. How can Social VR support larger teams, collectives, or organizational structures? How should multi-user work be managed and distributed? How persistent can these collaborations be after days, weeks, or months? CSCW research has long examined how group size alters coordination, governance, and knowledge sharing \cite{harris2019}. Yet, these insights are largely absent from current Social VR research. More empirical evidence is needed, including longitudinal studies, field deployments, and diverse samples. In particular, characterizing VR’s distinct affordances---such as spatial partitioning, altered scales and distances, and subgroup interactions---can provide new possibilities to enable collectives that are constrained by physical spaces and require embodied interactions. 

Second, most studies have relied on convenience samples---often college students and Western contexts---which limits the generalizability of their findings. This raises broader concerns that many Social VR systems may be implicitly designed around Western norms and expectations, potentially overlooking cultural differences in interaction styles, privacy preferences, or embodiment needs that could influence how people from diverse backgrounds experience and benefit from Social VR. Future research should recruit broader participant pools, including non-student populations, individuals with limited exposure to VR technologies, marginalized groups, and participants from varied cultural contexts. Broadening these samples will help diversify and strengthen the external validity of Social VR research.

Third, the review also highlights the lack of a robust ecosystem of applications in Social VR specifically designed for collaboration. Most studies relied on general-purpose platforms (e.g., AltspaceVR, VRChat), and some researchers tested them for specific collaborative purposes. While there is variety in productivity applications (e.g., Slack, MS Teams, Gmail), Social VR lacks dedicated, interoperable tools designed to integrate into everyday collaborative workflows. As suggested by our themes, Social VR requires not only designing the productivity features and tools to enable work but also considering how users experience collaboration contingent on individual, team, and environmental factors.

The corpus analysis also revealed that ethics and accessibility require consistent analysis, reporting, and discussion. HMD-based systems present significant barriers for users with visual, auditory, or motor impairments, raising concerns about adaptability and inclusion. Moreover, more discussions are needed to address ethical challenges, such as surveillance, harassment, and biometric data collection (e.g., motion, eye data). Many of these papers rarely specified what signals were recorded, for how long they were retained, what protections were in place, or whether accommodation features were available. Future studies should make these design choices, data collection practices, and accessibility features explicit. Accessibility is essential to prevent Social VR from becoming exclusionary.  

These mixed findings raise the broader question of ``\textit{why Social VR?}'' That is, under what conditions does it provide advantages over video calls or in-person meetings, and when does it fall short? Based on this corpus, the clearest advantages emerge in tasks that require room-scale layout, shared 3D manipulation, joint navigation, or view sharing---capabilities \emph{beyond} standard video conferencing. Many studies compared Social VR to video calls, with few comparing to in-person collaboration. As such, the current research leaves limited evidence about how these modes differ and how Social VR can reconfigure face-to-face dynamics. Overall, the studies indicate that Social VR is most effective when the collaborative task depends on embodied and spatial interaction. However, many applications continue to attempt to directly mimic physical offices or simply enable conversational dynamics, leading to increased frustration and costs compared to existing solutions.

\subsection{Design Implications}
Together, these themes reveal design conditions shaping Social VR collaboration: how expressive embodied interaction shapes individual experience, how groups coordinate and build trust, and how environments support joint action. Building on these thematic insights, we identify three design implications that translate these patterns into actionable directions for future systems.

\paragraph{Prioritizing Interaction Fidelity over Visual Fidelity}
Across the corpus, breakdowns in presence and social presence are frequently caused by low-fidelity motion, delayed or unnatural hand alignment, missing or awkward expressive cues, and limited support for natural gestures. These shortcomings hindered both the sense of ``being there'' and of ``being there with others,'' reducing engagement and limiting the authenticity of collaborative exchanges described in the sub-themes. Improving core interaction fidelity---e.g., stable object manipulation, predictable locomotion, low-latency input, local interpolation of movement, and expressive baseline cues when eye or face tracking is unavailable---should be a priority rather than just making the avatars more ``realistic.'' Collaboration benefits more from responsive embodied action than from avatar photorealism. Future systems should invest in reliable and expressive interaction as the foundation for collaborative behaviors in Social VR. 

\paragraph{Centering Collaboration around Shared Objects, Tools, and Spatial Coordination Mechanisms} 
Across the studies, collaboration improved when groups could easily create, manipulate, and reference shared artifacts. When tools were rigid, limited, or difficult to access, teams reported reduced cohesion, situational awareness, and engagement. The themes demonstrated how virtual objects serve not only as functional resources but also as social anchors, enabling collaborators to maintain a mutual focus and coordinate their actions. Additionally, spatial affordances---such as view-sharing, group-preserving teleportation, and visibility-balanced layouts---played a major role in enabling common ground. Thus, Social VR systems should emphasize robust, low-friction multi-user tools, persistent artifacts, and spatial mechanisms that enhance group awareness. These features support collaborative fluency and reduce ``re-entry costs'' that appear when teams must re-establish context.

\paragraph{Designing for Trust, Safety, Accessibility, and Transparency}
Findings across themes highlight that trust, safety, and comfort strongly influence participation and sustained collaboration. Privacy violations, proximity discomfort, unpredictable avatar movements, and inaccessible interaction modes eroded trust and undermined users' sense of social ease. Similarly, missing accommodations---such as captions, alternative input options, or personal-space controls---limited inclusivity and led to uneven participation. A design implication is that collaborative effectiveness depends on whether users feel protected, respected, and able to interact on their own terms. Incorporating accommodation elements should be a design priority in future systems. Moreover, many articles from the corpus provided limited detail about behavioral signal collection and data handling. Transparent reporting to the users of what signals are captured, how long they are retained, and what protections are applied can improve trust. Embedding safety, accessibility, and transparency directly into system design will broaden participation and help ensure Social VR remains comfortable and sustainable. 

\subsection{Limitations and Future Work}
It is important to acknowledge the limitations of this paper. First, we focused on peer-reviewed venues in the ACM Digital Library, IEEE Xplore, and Web of Science, core outlets for CSCW, HCI, and Social VR research. This ensured a reliable corpus but excluded pre-prints, dissertations, reports, and other gray literature. databases (e.g., Scopus, arXiv) may provide wider coverage. Future reviews could include non-peer-reviewed work---with explicit screening procedures---as industry-influences VR development.

Second, we limited our corpus to the period from 2004 to 2024 to capture the era of HMD-based Social VR. Earlier research on CVEs, CAVEs, and other screen-based platforms were excluded. While this decision allowed us to concentrate on the recent technologies shaping the current Social VR ecosystem, future reviews could revisit how pre-2004 work informs current systems and identify which insights have been lost or should be re-studied.

Third, our focus was on Social VR, defined as multi-user experiences mediated through HMDs. Other immersive technologies, such as Augmented Reality (AR) or Mixed Reality (MR), were excluded from our review. This maintained conceptual clarity in treating VR as a distinct medium. However, this restricts generalizability to new Extended Reality (XR) modalities. As the XR ecosystem becomes more widespread, future work should investigate how this broad spectrum of immersive technologies supports collaboration, potentially identifying new affordances and frameworks.

Fourth, our reliance on collaboration and work-related search terms may have excluded studies on Social VR in specific contexts, such as with older adults or marginalized groups, but did not explicitly frame these as ``collaboration.'' Future reviews should broaden collaboration definitions to include contexts where individuals pursue shared goals.

Fifth, scoping reviews are designed to map breadth rather than evaluate evidence strength. Because the focus is on coverage rather than depth, the synthesis remains descriptive, identifying patterns across studies without assessing reliability of individual findings or comparing effect sizes. Future systematic reviews or meta-analyses could build on this to rigorously assess evidence quality and consistency.

Lastly, screening and thematic coding were led by one researcher, with decisions discussed in meetings. While having meetings increased the consistency and robustness of these themes, the emerging codes depended on the first researcher, introducing bias in the analysis. Future studies should adopt multi-coder approaches or other assisted tools to prevent bias and avoid missing relevant codes.

\section{Conclusion}
This scoping review provides a comprehensive analysis of how Social VR has been studied as a medium for collaboration. Drawing from 62 relevant articles, we identified three conceptual levels (i.e., individual, team, and affordances) that shape how collaboration unfolds in Social VR. Our findings highlight its potential to enhance teamwork by fostering presence, supporting expressive and customizable avatars, and enabling social presence through nonverbal and spatial cues. The review also revealed important gaps in this research area, such as examining long-term collaboration dynamics, team-level processes such as trust-building, and the influence of avatar design on shaping self-presentation and group identity. 

Building on these insights, this paper encourages researchers and practitioners to move beyond short-term experimental designs and explore how Social VR can support sustained collaboration across diverse teams, domains, and cultural contexts. Taken together, this review highlights both the promise and the limitations of Social VR as a medium for facilitating collaboration. For CSCW, it highlights a renewed opportunity to shape the theories, methods, and design practices that will determine whether Social VR becomes a fleeting novelty or a sustainable foundation for the future of remote collaboration.

\begin{acks}
This work was supported by the Alfred P. Sloan Foundation (G-2024-22427).
\end{acks}

\bibliographystyle{ACM-Reference-Format}

\bibliography{References}
% \printbibliography

\received{May 2025}
\received[revised]{November 2025}
\received[accepted]{December 2025}

\clearpage
\appendix
\setcounter{table}{0}

\renewcommand{\thetable}{A\arabic{table}}
%TC:ignore
\section{List of included articles in the final corpus}
\label{appendix:corpus-table}

Overview of the studies included in the final corpus, along with their corresponding themes and subthemes.

\renewcommand{\arraystretch}{1.2}
{\footnotesize
\begin{longtable}{p{0.02cm} p{7.5cm} p{0.5cm} p{3.5cm}}

\caption{Overview of the final Social VR collaboration corpus.}
\Description{Long table listing all articles included in the final Social Virtual Reality collaboration corpus. Each row reports an article identifier, title, reference citation, and the sub-themes to which the article was assigned.}
\label{tab:final-corpus} \\
\toprule
\textbf{ID} & \textbf{Title} & \textbf{Ref} & \textbf{Themes} \\
\midrule
\endfirsthead

\toprule
\textbf{ID} & \textbf{Title} & \textbf{Ref} & \textbf{Themes} \\
\midrule
\endhead

\bottomrule
\endfoot

    1 & Conveying perspective in multi-user virtual reality collaborations & \cite{freiwald2020conveying} & Realistic \& Functional Environments \\

    2 & CakeVR: A Social Virtual Reality (VR) Tool for Co-designing Cakes & \cite{mei2021cakevr} & Sense of Presence; Social Presence \\
    
    3 & Falling Asleep Together: What Makes Activities in Social Virtual Reality Meaningful to Users & \cite{maloney2020falling} & Real Collaboration \\

    4 & Enhanced Auditoriums for Attending Talks in Social Virtual Reality & \cite{weissker2023enhanced} & Social Presence \\

    5 & Just undo it: exploring undo mechanics in multi-user virtual reality & \cite{rasch2024just} & Virtual Tools \& Objects \\

    6 & Remote and Collaborative Virtual Reality Experiments via Social VR Platforms & \cite{saffo2021remote} & Realistic \& Functional Environments \\

    7 & Communication in Immersive Social Virtual Reality: A Systematic Review of 10 Years’ Studies & \cite{wei2022communication} & Social Presence\\

    8 & Working Together Apart through Embodiment: Engaging in Everyday Collaborative Activities in Social Virtual Reality & \cite{freeman2022working} & Realistic \& Functional Environments \\

    9 & Give-Me-A-Hand: The Effect of Partner’s Gender on Collaboration Quality in Virtual Reality & \cite{yassien2021give} & Avatar Design \\

    10 & Enhancing collaborative manipulation through the use of feedback and awareness in CVEs & \cite{garcia2008enhancing} & Virtual Tools \& Objects \\

    11 & A collaborative VR visualization environment for offshore engineering projects & \cite{dos2011collaborative} & Virtual Tools \& Objects \\

    12 & Effect of gender on immersion in collaborative iODV applications & \cite{kallioniemi2017effect} & Avatar Design \\

    13 & A review on effective closely-coupled collaboration using immersive CVE's & \cite{otto2006review} & Virtual Tools \& Objects \\

    14 & Subjective evaluation of group user QoE in collaborative virtual environment (CVE) & \cite{moharana2022subjective} & Social Presence \\

    15 & Cone of Vision as a Behavioural Cue for VR Collaboration & \cite{bovo2022cone} & Social Presence \\

    16 & Inter-brain Synchrony and Eye Gaze Direction During Collaboration in VR & \cite{gumilar2022inter} & Social Engagement \\

    17 & Impact of Augmented Engagement Model for Collaborative Avatars on a Collaborative Task in Virtual Reality & \cite{le2024impact} & Social Engagement \\

    18 & Visual deictic reference in a collaborative virtual environment & \cite{duchowski2004visual} &Social Engagement \\

    19 & A Design Space for Social Presence in VR & \cite{yassien2020design} & Avatar Design; Teammate Trust \\

    20 & VRGit: A Version Control System for Collaborative Content Creation in Virtual Reality & \cite{zhang2023vrgit} &Realistic \& Functional Environments \\

    21 & blue-c API: a multimedia and 3D video enhanced toolkit for collaborative VR and telepresence & \cite{naef2004blue} & Spatial Audio \& Movement \\

    22 & Let’s Face It: Influence of Facial Expressions on Social Presence in Collaborative Virtual Reality & \cite{kimmel2023let} & Social Presence \\

    23 & Identifying Cognitive and Creative Support Needs for Remote Scientific Collaboration using VR: Practices, Affordances, and Design Implications & \cite{olaosebikan2022identifying} & Sense of Presence; Social Presence \\

    24 & Towards Immersive Collaborative Sensemaking & \cite{yang2022towards} & Social Engagement \\

    25 & Designing Virtual Environments for Social Engagement in Older Adults: A Qualitative Multi-site Study & \cite{xu2023designing} & Social Engagement \\

    26 & Users' locomotor behavior in collaborative virtual reality & \cite{rios2018users} & Spatial Audio \& Movement \\

    27 & Invoke: A Collaborative Virtual Reality Tool for Spatial Audio Production Using Voice-Based Trajectory Sketching & \cite{deacon2023invoke} & Spatial Audio \& Movement \\

    28 & LeMo: Exploring Virtual Space for Collaborative Creativity & \cite{men2019lemo} & Realistic \& Functional Environments \\

    29 & Being Social in VR Meetings: A Landscape Analysis of Current Tools & \cite{osborne2023being} & Avatar Design; Real Collaboration; Realistic \& Functional Environments; Virtual Tools \& Objects  \\

    30 & AMMP-Vis: a collaborative virtual environment for molecular modeling & \cite{chastine2005ammp} & Social Presence \\

    31 & CollaVR: Collaborative In-Headset Review for VR Video & \cite{nguyen2017collavr} & Virtual Tools \& Objects \\

    32 & Synchrony within Triads using Virtual Reality & \cite{miller2021synchrony} & Sense of Presence \\

    33 & Sushi with Einstein: Enhancing Hybrid Live Events with LLM-Based Virtual Humans & \cite{shoa2023sushi} & Avatar Design \\

    34 & Social Simon Effect in Virtual Reality: Investigating the Impact of Co-actor Avatar's Visual Representation & \cite{li2024social} & Avatar Design \\

    35 & Supporting social human communication between distributed walk-in displays & \cite{roberts2004supporting} & Virtual Tools \& Objects \\

    36 & Going, Going, Gone: Exploring Intention Communication for Multi-User Locomotion in Virtual Reality & \cite{rasch2023going} & Spatial Audio \& Movement \\

    37 & Effects of Avatar’s Symbolic Gesture in Virtual Reality Brainstorming & \cite{ide2020effects} & Social Presence \\

    38 & Proof-of-Concept Study to Evaluate the Impact of Spatial Audio on Social Presence and User Behavior in Multi-Modal VR Communication & \cite{immohr2023proof} & Spatial Audio \& Movement \\

    39 & The Effects of Self- and External Perception of Avatars on Cognitive Task Performance in Virtual Reality & \cite{kocur2020effects} & Avatar Design \\

    40 & Dyadic interactions with avatars in immersive virtual environments: high fiving & \cite{young2015dyadic} & Avatar Design \\

    41 & Lessons From Working in the Metaverse: Challenges, Choices, and Implications from a Case Study & \cite{park2024lessons} & Real Collaboration\\

    42 & Design principles for social exchange in social virtual reality-enabled virtual teams & \cite{torro2023design} & Avatar Design; Real Collaboration; Teammate Trust \\

    43 & Studying the effect of symmetry in team structures on collaborative tasks in virtual reality & \cite{agnes2023studying} & Social Presence \\

    44 & Working Together on Diverse Tasks: A Longitudinal Study on Individual Workload, Presence and Emotional Recognition in Collaborative Virtual Environments & \cite{khojasteh2021working} &Sense of Presence \\

    45 & Communicating Information in Virtual Reality: Objectively Measuring Team Performance & \cite{moore2019communicating} & Social Engagement \\

    46 & Getting There Together: Group Navigation in Distributed Virtual Environments & \cite{tim2020getting} & Spatial Audio \& Movement \\

    47 & Supporting a closely coupled task between a distributed team: Using immersive virtual reality technology & \cite{roberts2005supporting} & Virtual Tools \& Objects \\

    48 & Group Navigation for Guided Tours in Distributed Virtual Environments & \cite{weissker2021group} & Realistic \& Functional Environments \\

    49 & Collaborative virtual reality platform for visualizing space data and mission planning & \cite{garcia2019collaborative} & Virtual Tools \& Objects \\

    50 & See-through techniques for referential awareness in collaborative virtual reality & \cite{argelaguet2011see} & Virtual Tools \& Objects \\

    51 & Measuring the collaboration degree in immersive 3D collaborative virtual environments & \cite{sarmiento2014measuring} & Spatial Audio \& Movement \\

    52 & Multi-Ray Jumping: Comprehensible Group Navigation for Collocated Users in Immersive Virtual Reality & \cite{weissker2019multi} & Spatial Audio \& Movement \\

    53 & Multiparty Holomeetings: Toward a New Era of Low-Cost Volumetric Holographic Meetings in Virtual Reality & \cite{langa2022multiparty} & Realistic \& Functional Environments \\

    54 & A Longitudinal Study of Task Performance, Head Movements, Subjective Report, Simulator Sickness, and Transformed Social Interaction in Collaborative Virtual Environments & \cite{bailenson2006longitudinal} & Real Collaboration \\

    55 & Evaluating the Effect of Binaural Auralization on Audiovisual Plausibility and Communication Behavior in Virtual Reality & \cite{immohr2024evaluating} & Spatial Audio \& Movement \\

    56 & Physiological Synchrony in a Collaborative Virtual Reality Task & \cite{moharana2023physiological} & Social Engagement \\

    57 & Subjective Evaluation of the Impact of Spatial Audio on Triadic Communication in Virtual Reality & \cite{immohr2024subjective} & Spatial Audio \& Movement \\

    58 & Kinetic Connections: Exploring the Impact of Realistic Body Movements on Social Presence in Collaborative Virtual Reality & \cite{kimmel2024kinetic} & Avatar Design \\

    59 & Hands or Controllers? How Input Devices and Audio Impact Collaborative Virtual Reality & \cite{adkins2024hands} & Social Presence \\

    60 & GazeMolVR: Sharing Eye-Gaze Cues in a Collaborative VR Environment for Molecular Visualization & \cite{darbar2024gazemolvr} & Social Presence \\

    61 & Social virtual reality: systematic review of virtual teamwork with head-mounted displays & \cite{dey2024social} & Social Presence; Real Collaboration; Virtual Tools \& Objects \\

    62 & Exploring sex differences in collaborative virtual environments for participation equality and user experience & \cite{yang2024exploring} & Avatar Design \\

    \bottomrule
  \end{longtable}
}

%TC:endignore
\end{document}